\documentclass[11pt]{book}
\usepackage{Wiley-AuthoringTemplate}
\usepackage[authoryear]{natbib}% for name-date citation comment the below line
\makeindex% for index generation

\begin{document}

\frontmatter

\mainmatter
%%%%%%%%%%%%%%%%%%%%%%%%%%%%%%%%%%

\def\solphys{Sol. Phys.}
\def\aap{A\&A}
\def\aa{Astron. Astrophys.}
\def\aapr{Astro. \& Astrophys. Review}
\def\ag{Ann. Geophys.}
\def\apj{Astrophys. J.}
\def\apjl{Astrophysics J. Lett.}
\def\apjsup{Astrophys. J. Suppl. Ser.}
\def\araa{Ann. Review of Astron. and Astrophy.}
\def\cjp{Can. J. Phys.}
\def\cpc{Comput. Phys. Commun.}
\def\grl{Geophys. Res. Lett.}
\def\jatp{J. Atmos. Terr. Phys.}
\def\jcp{J. Comput. Phys.}
\def\jgg{J. Geomagn. Geoelectr.}
\def\jgr{J. Geophys. Res.}
\def\mnras{Mon. Not. R. Astron. Soc.}
\def\pasj{Publ. Astron. Soc. Jap.}
\def\pf{Phys. Fluids}
\def\pfb{Phys. Fluids B}
\def\ppcf{Plasma Phys. Contr. Fusion}
\def\plas{Plasma Phys.}
\def\prl{Phys. Rev. Lett.}
\def\ps{Phys. Scr.}
\def\pss{Planet. Space Sci.}
\def\rgsp{Rev. Geophys.}
\def\solphys{Solar Physics}
\def\sovast{Soviet Astronomy}
\def\sp{Solar Physics}
\def\ssp{Space Sci. Rev.}
\def\asr{Adv. Space Res.}
\def\physrep{Phys. Rep.}%##说明：若是用beamer则上一应为：\documentclass[CJK, 12pt]{beamer} 一写要加CJK，否则出错
\newcommand{\yprl}[3]{ #1, {PhRvL,} {#2}, #3}

\chapter[Observations of Magnetic Helicity Proxies in Solar Photosphere: Helicity with Solar Cycles]{Observations of Magnetic Helicity Proxies in Solar Photosphere:  Helicity with Solar Cycles}

\author*{Hongqi Zhang*{$^{1}$}}
\author{Shangbin Yang{$^{1,2}$}}
\author{Haiqing Xu{$^{1}$}}
\author{Xiao Yang{$^{1}$}}
\author{Jie Chen{$^{1}$}}
\author{Jihong Liu$^{1,3}$}

\address[1]{\orgdiv{National Astronomical Observatories},
\orgname{Chinese Academy of Sciences},
\postcode{100101}, \countrypart{ Chaoyang District},
    \city{ Beijing}, \street{20A Datun Road}, \country{China}}

\address[2]{\orgdiv{University of Chinese Academy of Sciences},
\postcode{100049},
     \city{ Beijing}, \street{19A Yuquan Road}, \country{China} }%

\address[3]{\orgdiv{Shijiazhuang University },
\postcode{050035},
     \city{ Shijiazhuang}, \street{288A Zhufeng Road}, \country{China}    }%

\address*{Corresponding Author: Hongqi Zhang; \email{hzhang@bao.ac.cn}}
%%%%%%%%%%%%%%%%%%%%%%%%%%%%%%%%%

\maketitle% This tag is required to print author and address in the output

\begin{abstract}{Abstract}
{%\color{red}  
 Observations of magnetic helicity transportation through the solar photosphere reflect the interaction of turbulent plasma movements and magnetic fields in the solar dynamo process. In this chapter,  we have reviewed the research process of magnetic helicity inferred from the observed solar magnetic fields in the photosphere and also the solar morphological configurations with solar cycles. After introducing some achievements in the study of magnetic helicity, some key points would like to be summarized. 
 
The magnetic (current) helicity in the solar surface layer presents a statistical distribution similar to that of the sunspot butterfly diagram, but its maximum value is delayed from the extreme value of the sunspot butterfly diagram and corresponds in the phase with the statistical eruption of solar flares. During the spatial transport of magnetic (current) helicity from the interior of the sun into the interplanetary space at the time-space scale of the solar cycle, it shows the statistical distribution and the fluctuation with the hemispheric sign rule. These show that the current helicity and magnetic helicity transport calculation methods are complementary to each other.

We also notice that the study of the inherent relationship between magnetic helicity and the solar cycle still depends on the observed accuracy of the solar magnetic field.
}
\end{abstract}


\section{Introduction}

%{\color{red}
The study of solar magnetic helicity with the solar cycles is an interesting topic. The solar activity cycle is first found from the observation results of sunspots, which have 11-year cycles \citep{Schwabe43}. Observation of solar magnetic fields with magnetographs has brought human cognition for understanding solar activities to a new stage \citep{Hale08}. 

A large number of observations show that the solar active regions are the place where the magnetic field energy on the solar surface is most concentrated \citep[cf.][]{Fisher00,Schmieder14,Toriumi20}. Accompanied by flares or coronal mass ejections, the energy of the magnetic field is transported from the solar subatmosphere into the interplanetary space - strong solar storms \citep[cf.][]{WangXu02,Aschwanden05}. It is found that the intense solar eruptive phenomena are often connected with the strong magnetic helicity \citep[cf.][]{Low1996,BothmerSchwenn98,Yurchyshyn01,Nindos02,Lynch05,TorokKliem05}.

Due to the opacity of the solar atmosphere, that is, even using the method of helioseismology to diagnose the structure inside the sun,  it is still little known about the situation of the magnetic fields inside of the sun in detail \citep{Parker1979CMF,Fan09}. The magnetic field distribution on the surface of the sun brings important information about the formation and transport of magnetic field energy with helicity in the subatmophere.

People try to explore the mechanism of the formation of the magnetic field inside of the sun through the theory of the solar dynamo, to explain the statistical phenomenon that the solar activity presents an 11-year cycle, and so on  \citep[cf.][]{par55,M78,krarad80,ZRS83}. The usual theories believe that the chilarity (helicity) of the magnetic field is also one of the important pieces of evidence for the formation of the magnetic field inside the sun, and believe that the helicity tends to show opposite signs in both solar hemispheres in the process of solar dynamos \citep[cf.][]{Seehafer94}. It is normally believed that the observed magnetic helicity in the solar surface brings the message of the $\alpha$-effect in the deep solar convection zone from the point of view of the mean-field $\alpha\omega$ dynamo model \citep[cf.][]{bra-sub:05} or also the flux-transport model  \citep[cf.][]{Choudhuri04}.    

The statistical analysis of the observed helical magnetic field in the solar atmosphere, such as in the solar active regions, provides a window for studying the possible formation mechanism of the solar active cycles with the solar dynamos. 

In the following, we present the current helicity in active regions, the injection of the magnetic helicity with solar cycles inferred from observed photospheric magnetic fields, and the relationship with solar eruptive activities. We also discuss some questions from the observations of solar magnetic helicity.

\section{Magnetic Helicity from Observations}

\subsection{Early Morphological Observations of helical Patterns in Solar atmosphere}
\begin{figure}[!h]
\begin{center}
%\epsscale{1.0}
\includegraphics[width=80mm]{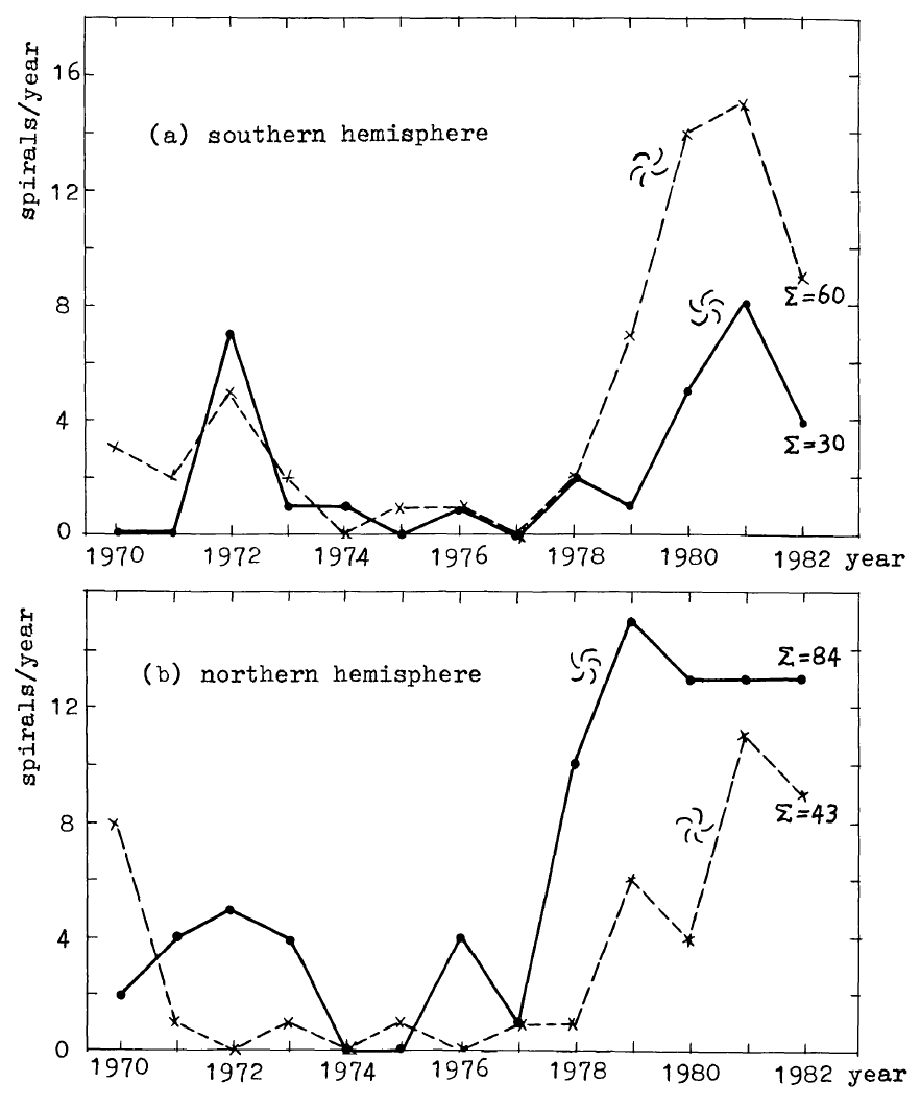}
%\vspace*{1mm}
\end{center}
\caption{Distribution of spiral patterns of active regions in solar southern (a) and northern (b) hemispheres. The ordinates (spirals/year) are the statistical number of twisted sunspots per year. The solid (dashed) lines indicate the twist of sunspots in right (left) handedness.
 After \cite{Ding87}
}\label{fig:Ding1987fig-3}
\end{figure}

%}
The chiral patterns of active regions on the Sun were firstly found by \cite{Hale08,Richardson41} who found that most of the sunspots (about $80\%$)  with discernible whirls had phenomenological counterclockwise (clockwise) rotations of the fibril patterns in the northern (southern) hemisphere from the observations. It has been further confirmed and statistically presented on the distribution of helical sunspots in both hemispheres in 1970-1982 by \cite{Ding87}, which is covered about a solar cycle, as shown in Figure \ref{fig:Ding1987fig-3}.  Interestingly, only about 66\% of sunspots follow the helical sign rule.

\begin{figure}[!h]%[H]
\begin{center}
\includegraphics[width=70mm]{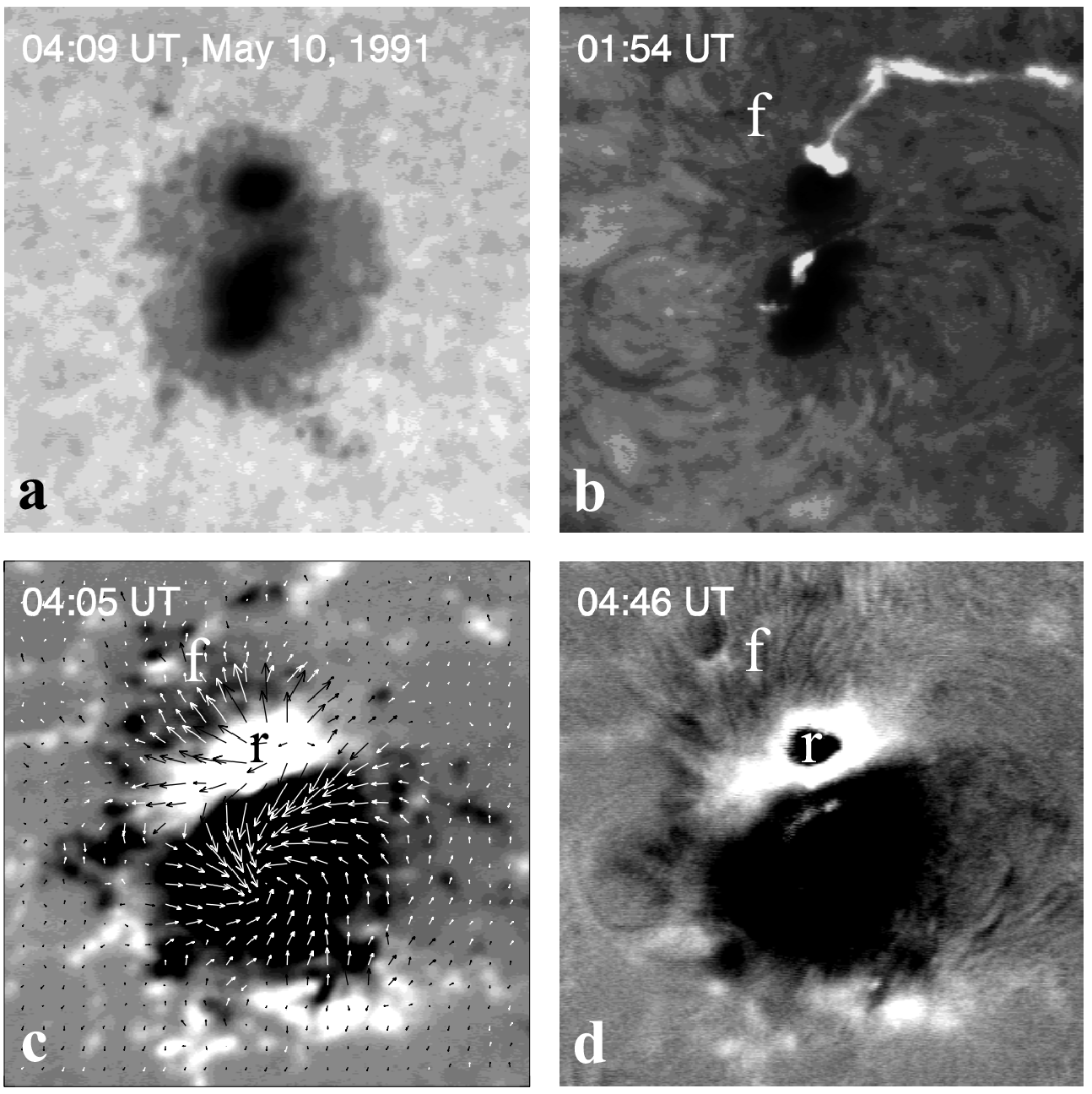}
\end{center}
\caption{Active region NOAA 6619 observed on May 10, 1991. Top left: Photospheric filtergram. Bottom left: Photospheric vector magnetogram. Top right: H$\beta$ filtergram. Bottom right: longitudinal magnetogram in Chromosphere from  H$\beta$. Black (white) is negative (positive) polarity in the magnetograms. %The size for each figure is $2.'8\times 2.'8$. 
After \cite{Zhang19}.
\label{fig:ar6619}}
\end{figure}

The observations of magnetic fields in the solar photosphere using solar magnetographs or Stokes polarimeters are a general method to diagnose the twisted pattern of the field in sunspot regions.
As an example, Figure \ref{fig:ar6619} shows the photospheric vector and chromospheric longitudinal magnetic field in a delta active region  \citep{Zhang19,Zhang20}.  It is found that the horizontal twisting component of the photospheric vector magnetic field of the active region is consistent with the features of the H$\beta$ chromospheric magnetic field and also the chromospheric morphological features. These confirm that the morphological pattern of spiral spot features as before observed by \cite{Hale08,Richardson41,Ding87} reflects the configuration of the magnetic fields in the atmosphere (such as marked by $f$ in Figure \ref{fig:ar6619}).

Moreover, the evidence on the hemispheric sign rule of the helical features in the chromosphere was also indicated by \cite{Martin94}. They found morphologically that the majority of quiescent filaments were dextral/sinistral in the northern/southern hemisphere. \cite{Rust94} pointed out the relationship amount the patterns of chirality in sunspots, active regions, filaments, and interplanetary magnetic clouds.  These imply that the helical patterns on the Sun relate to the different scale magnetic fields, due to the dominant controls of the magnetic fields in the sunspots of active regions, filaments, clouds, and so on. This means that the helical features extend from the photosphere into the high atmosphere and the interplanetary space.
It is noticed that the low sensitivity of chromospheric spectral lines to magnetic field observations, non-local thermal dynamic equilibrium, and disturbance of photospheric blended spectral lines (such as in the wing of H$\beta$ line) affect the measurement of the solar chromospheric magnetic field, and make the difficulty to obtain the chromospheric vector magnetic field %and to derive the horizontal component of the electric current below the chromosphere 
\citep{Zhang19,Zhang20}. The reversal patch $r$ in the umbra of chromospheric H$\beta$ magnetogram in Figure \ref{fig:ar6619} is caused by the disturbance of photospheric blended spectral lines in the wing of the H$\beta$ line.

%{\color{red}
\subsection{Magnetic Helicity from Solar Vector Magnetic Fields}

The magnetic helicity is the volume integral   \citep{Woltjer58a,Woltjer58b}   
\begin{equation}
\label{heli}
H_m=\int_V{\bf A}\cdot \nabla \times {\bf A}d^3x,
\end{equation}
where $\bf A$ is the magnetic vector potential and is not an immediately observed quantity and does not satisfy the requirement of gauge invariance.  To simplify calculations as employ the Coulomb gauge \citep{moff69,Arnold74,BergerField84}
\begin{equation}
\label{eq:magneticpotenBF}
{\bf A(x)}=-\frac{1}{4\pi}\int d^3x'\frac{\bf r}{r^3}\times{\bf B(x')},
\end{equation}
which gives
\begin{equation}
\label{ }
H_m=-\frac{1}{4\pi}\int d^3x\int d^3x'{\bf B(x)}\cdot\left[\frac{\bf r}{r^3}\times{\bf B(x')}\right],
\end{equation}
with the relationship ${\bf B}=\nabla\times{\bf A}$ \citep{Low15}.

\begin{figure}[!h]
\vspace{7pt} \centering
\includegraphics[width=1.\textwidth,angle=0]{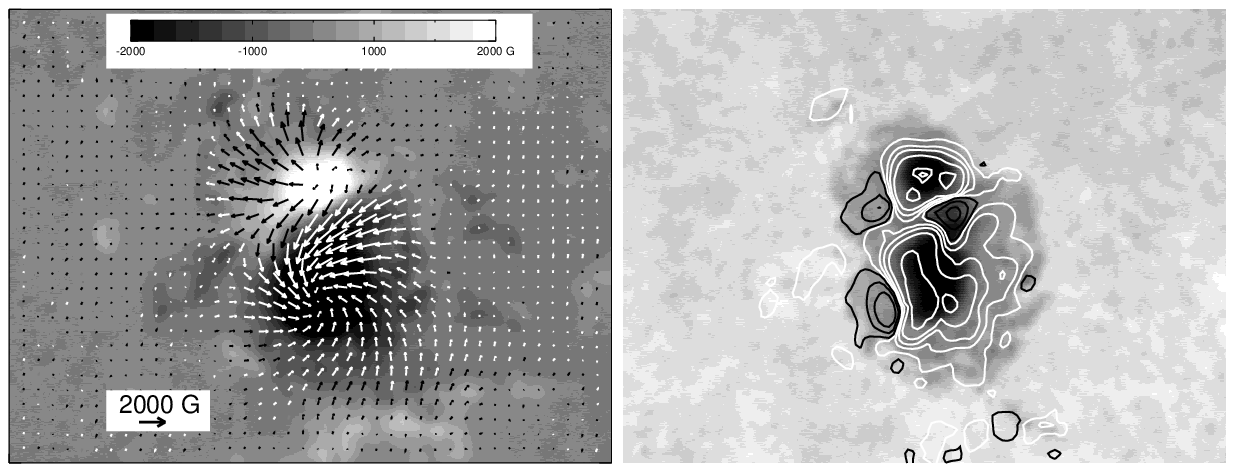}
\caption{
The active region NOAA 6619 was taken at Huairou Solar Observing Station on May 11, 1991,
at 03:26 UT. Left: the photospheric vector magnetogram.
Right: the electric current helicity density  $h_{cz}=({\bf B}\cdot\nabla\times {\bf B})_z$ (contours) overlapped by the filtergram of this active region;  the black (white) contours correspond to positive (negative) values of 0.01, 0.05, 0.1, 0.2, 0.5 G$^{2}$m$^{-1}$, respectively.
}
\label{fig:magcurrhelic}
\end{figure}

The current helicity $H_c$ is defined as
\begin{equation}
\label{eq:cuurhelicity}
H_c=\int_V{\bf B}\cdot\nabla\times {\bf B}d^3x,
\end{equation}
where $\bf B$ is the magnetic field and the current helicity density can be difined 
\begin{equation}
\label{ }
\begin{aligned}
   h_c &={\bf B}\cdot\nabla\times {\bf B}=({\bf B}\cdot\nabla\times {\bf B})_z+({\bf B}\cdot\nabla\times {\bf B})_t, 
  \end{aligned}
  \end{equation}
the subscript $z$ and $t$ mark the longitudinal and transverse components of current helicity respectively. The current helicity density $h_{cz}=({\bf B}\cdot\nabla\times {\bf B})_z$ is observable in the lower solar atmosphere by observed photospheric vector magnetogam in Figure \ref{fig:magcurrhelic}, while $({\bf B}\cdot\nabla\times {\bf B})_t$ is difficult  because it is impossible to achieve the construction of the real transverse component of electric current through the observations of the solar magnetic fields.

\begin{figure}[!h]%[H]
\begin{center}
\includegraphics[width=130mm]{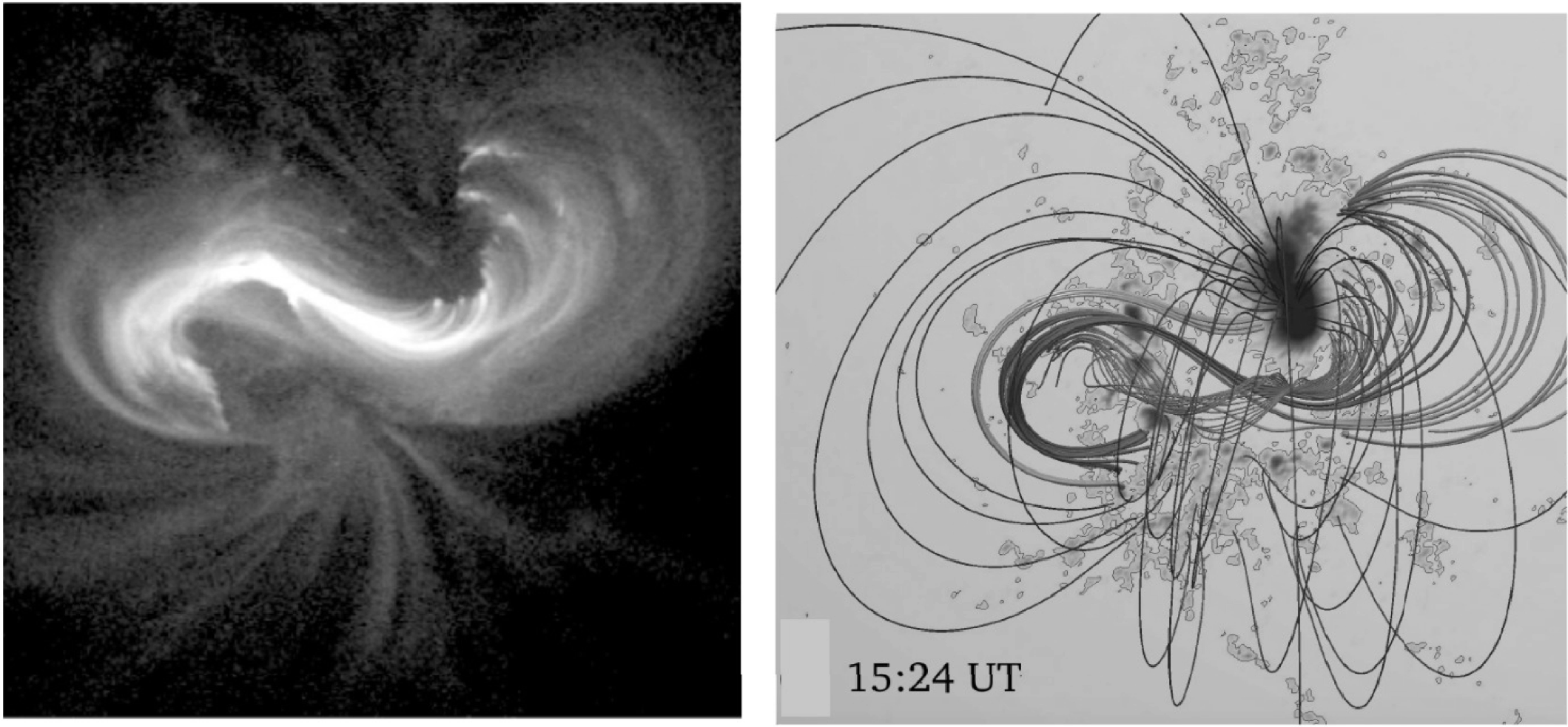}
\end{center}
\caption{Sigmoidal field lines wrapping around a flux rope in the active region 12158, which has a negative helicity (reverse S, left‐hand twist) on 10 September 2014, 2 hr before a flare. The left panel shows an observation in AIA 94 \AA; the right panel shows sigmoidal magnetic field lines using an NLFFF extrapolation based on the Grad‐Rubin method \citep{Gilchrist14}, overlying a  longitudinal component of photospheric magnetic field saturated at 2000 gauss. 
After \cite{Schmieder18}.
\label{fig:TopolSchAul}}
\end{figure}

The chirality of the magnetic structures can be analyzed morphologically due to the magnetic freezing effect of the plasma in the solar atmosphere. It reflects the spatial characteristics of the magnetic field helicity. This also expands the distribution and range of magnetic helicity in the solar atmosphere that we can analyze from an observable perspective.
Figure \ref{fig:TopolSchAul} shows the extrapolated nonlinear force free (NLFF) magnetic field calculated from the photospheric magnetic field in an active region to fit AIA 94\AA{} flare loops \citep{ZhaoJ16,Schmieder18}, which has a negative helicity.

From the above discussion, it is found that the distribution of the magnetic (current) helicity in the solar atmosphere and its evolution can be analyzed from different perspectives (the magnetic vector potential, magnetic field, and morphology). The key parameter is the magnetic field, especially from the observations of the photospheric magnetic field. 

%}

\section{Electric Current Helicity from Observations}

\subsection{Helicity from solar vector magnetic fields}

The early statstical research on the magnetic (current) helicity from the observations of solar magnetic fields was firstly presented by \cite{Seehafer90}, and a series of subsequent studies by \cite{Pevtsov95,Abramenko96,BZ98,Hagino05,Xu07}, etc.
%The current helicity $H_c$ is defined as
%\begin{equation}
%\label{eq:cuurhelicity}
%H_c=\int h_cdv=\int{\bf B}\cdot{\bf {{\bf {\nabla}} \times \,}} {\bf B}dv.
%\end{equation}
The observable current helicity density averaged over an active region is
\begin{equation}
\label{eq:meancurr}
\overline{h_{{\rm c}z}}=\overline{({\bf B}\cdot{\bf {{\bf {\nabla}} \times \,}} {\bf B})_{z}},
\end{equation}
which can be derived from the photospheric vector magnetograms (see Figure \ref{fig:magcurrhelic}).
In the approximation of local homogeneity and isotropy, the value of $h_{{\rm c}z}$ can be used to analyze the total current helicity  $h_{\rm c} $ \citep{Xu15}.
We need to notice that $h_{{\rm c}z}=1/3h_{\rm c} $ is an estimated value, due to the incompleteness for inversion of the solar magnetic field through the solar polarized spectral light \citep{Zhang19, Zhang20}.

Another proxy  of the magnetic field is twist, i.e., the force-free factor $\alpha={\bf B}\cdot( \nabla\times {\bf B})/B^2$, or an equivalent of the quantity averaged over an active region is the ratio 
\begin{equation}
\label{eq:meanalpha}
\alpha_{\rm av}=\overline{(\nabla\times {\bf B})_{z}/B_{z}}
\end{equation}
calculated from the observed vector magnetograms.

\cite{Pevtsov05} reported the result of a study of magnetic helicity in solar active regions in 1980-2000 (cycles 21-23) using $\alpha_{\rm av}$ as the proxy for current helicity. 
They did not see consistency between different instruments in regards to years of disobeying the hemispheric helicity rule.  
They found that the present data sets do not allow for making statistically significant inferences about the possible cyclic variation of the hemispheric helicity rule.
\cite{Xu07}  compared a series of vector magnetograms of the same active regions observed by different solar vector magnetographs. 
It is found that a similar tendency of calculated current helicities from the magnetograms was observed at the different observatories. Although there exist slight differences among them.

\begin{figure}[!h]
%\vskip -4mm
\centering
{\includegraphics[width=0.4\textwidth]{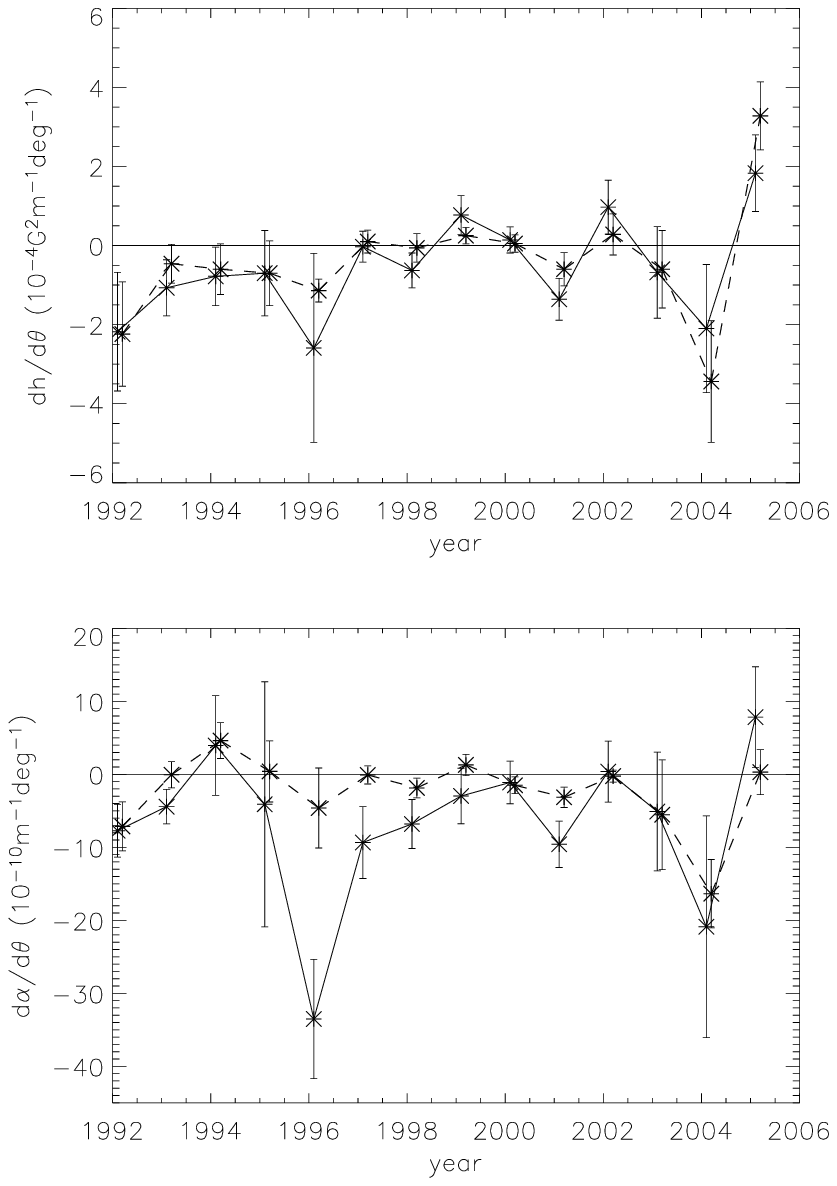}
\includegraphics[width=0.4\textwidth]{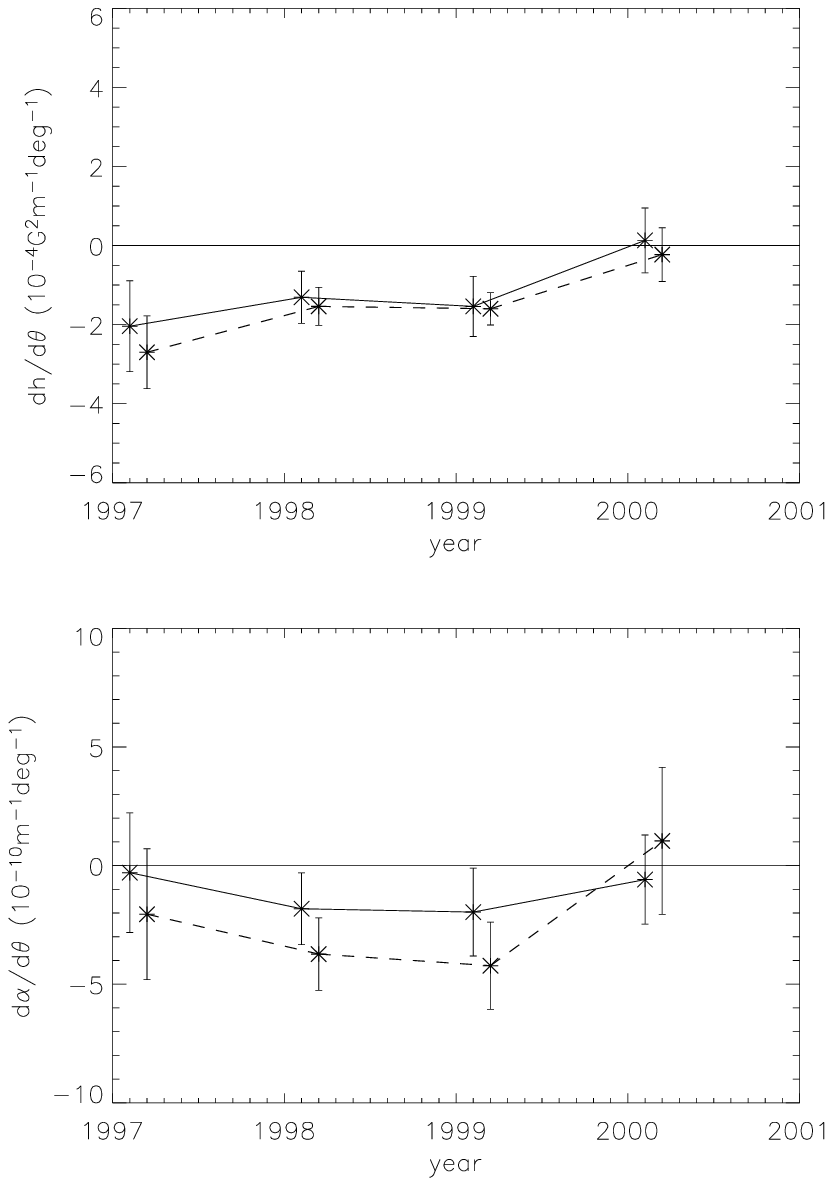}}
\caption{ Time variation of the mean slope of linear fit calculated from the latitudinal profile of helicity for each year over 1992--2005 (left) and 1997--2000 (right). The solid (dashed) lines in the left represent HR (MTK)  data, and in the right Mees (MTK)  data, respectively. The upper panel is \textit{d}$h_{\rm c}$/$d{\theta}$ and the lower is \textit{d}${\alpha}$$_{\rm ff}$/$d{\theta}$.  Error bars represent one sigma uncertainty of slope of the linear fit.   After \cite{Xu07} }
\label{fig:magnetogrps}
\end{figure}

The seemingly contradictory results of the helicity mentioned above may mainly come from the observations of the solar magnetic field, and the processing and analysis of the data. The diagnostic of solar magnetic fields based on these magnetically sensitive lines is accompanied by the approximations of the theoretical analysis, which probably brings the uncertainty of inversion of spectro-polarimetric observations to obtain magnetic fields.
It is important to prevent or avoid the influence of these errors and try to obtain relatively accurate measurements of solar magnetic fields if the defects of the different magnetographs can be ignored.
We probably can believe that more reasons probably bring the observed errors at the vector magnetographs from different observatories
\citep{Ai89,Ronan92,WangHM92,Sakuari95,Sakuari01,ZhangHQ03}  until now, as we wish to analyze in detail probably.

\subsection{Magnetic helicity with solar cycles}

The helicity evolution with the solar cycle is a notable topic. The mean distribution of photospheric current helicity density of active regions in the solar surface
by $\overline{h_{{\rm c}z}}=\overline{({\bf B}\cdot{\bf {{\bf {\nabla}} \times \,}} {\bf B})_{z}}$ in Eq. (\ref{eq:meancurr}) and also $\alpha_{{\rm av}}=\overline{(\nabla\times {\bf B})_{z}/B_{z}}$   in Eq. (\ref{eq:meanalpha}) in solar cycle 22 was statistically presented by \cite{BZ98} and \cite{ZhangBao98,ZhangBao99}, which are inferred from a series of photospheric vector magnetograms recorded at Huairou Solar Observing Station.

\begin{figure}[!h]
\centering
\includegraphics[angle=0,scale=.40]{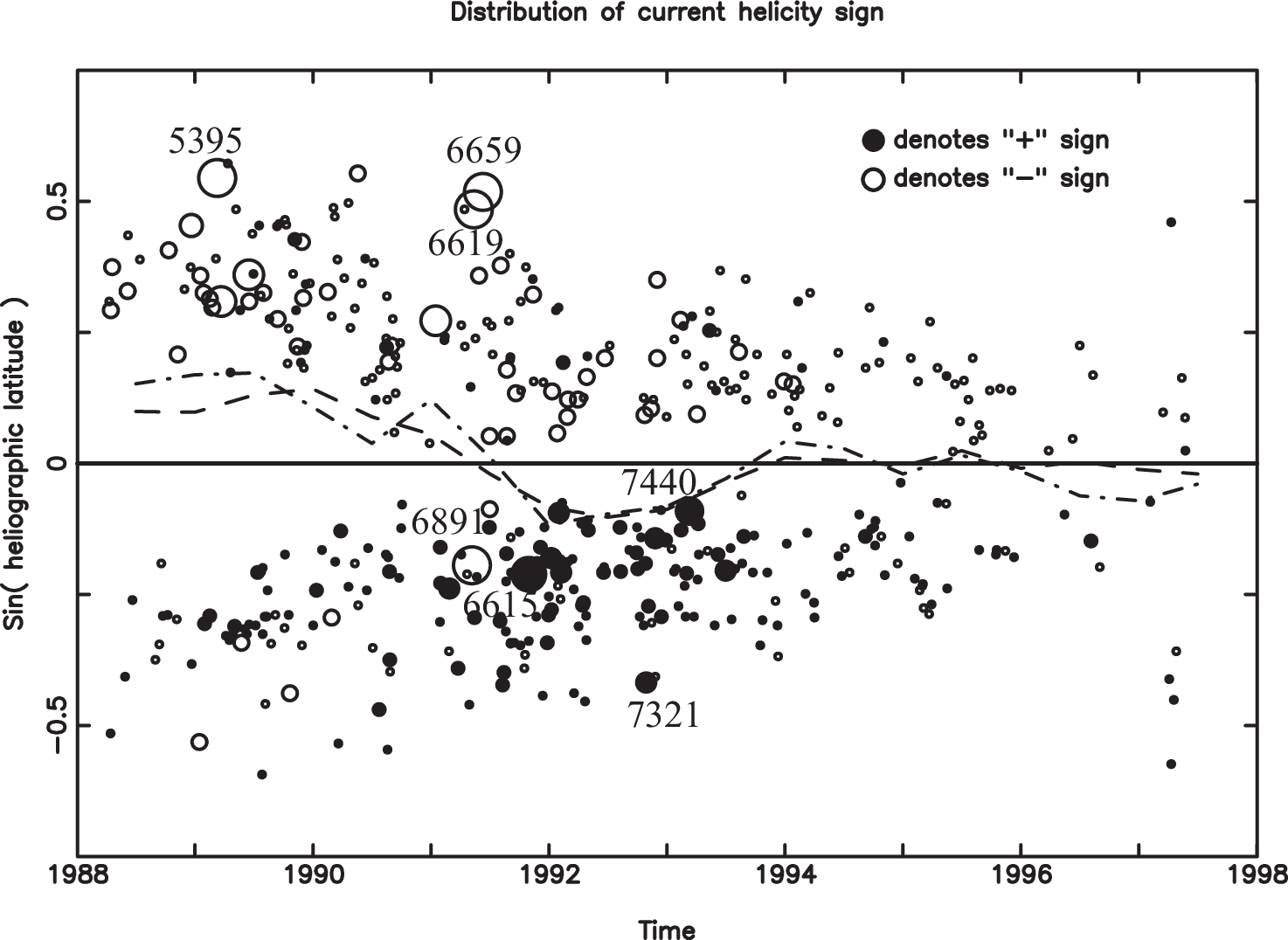}
\caption{Butterfly diagram of the mean current helicity density $h_{{\rm c}z}$. The sizes of the circles relate to the grades: 0, 1, 3, 5, 7 ($10^{-3}G^2m^{-1}$) of current helicity density. The white (black)  circles mark the negative (positive) sign of helicity. The dashed-dotted line presents the distributed average value of current helicity and the dashed line does the mean value of the imbalance of current helicity after the data smooth. Vertical coordinates represent the sine of latitude, and horizontal coordinates are expressed in years. From \cite{ZhangBao98}}\label{fig:zb_1998}
 \end{figure}

{%\color{red}
Figure \ref{fig:zb_1998} shows the distribution of the current helicity of active regions in the form of the Butterfly diagram observed at Huairou. It is found that besides most of the helicity of active regions following the hemispheric sign rule, some super $\delta$ active regions have been marked in the figure. Some high density of current helicity regions (such as NOAA 5395, 6659, and 6619) occurred at the high latitudes, while NOAA 7321 and 7440 occurred at the low ones. The notable is that the active region NOAA 6891 shows the reversal sign of helicity.
}

\begin{figure}[!h]
\centering
%\hbox{
%\vskip -58mm
\hskip -8mm
\includegraphics[width=110mm,angle=0]{%fig8_1.ps
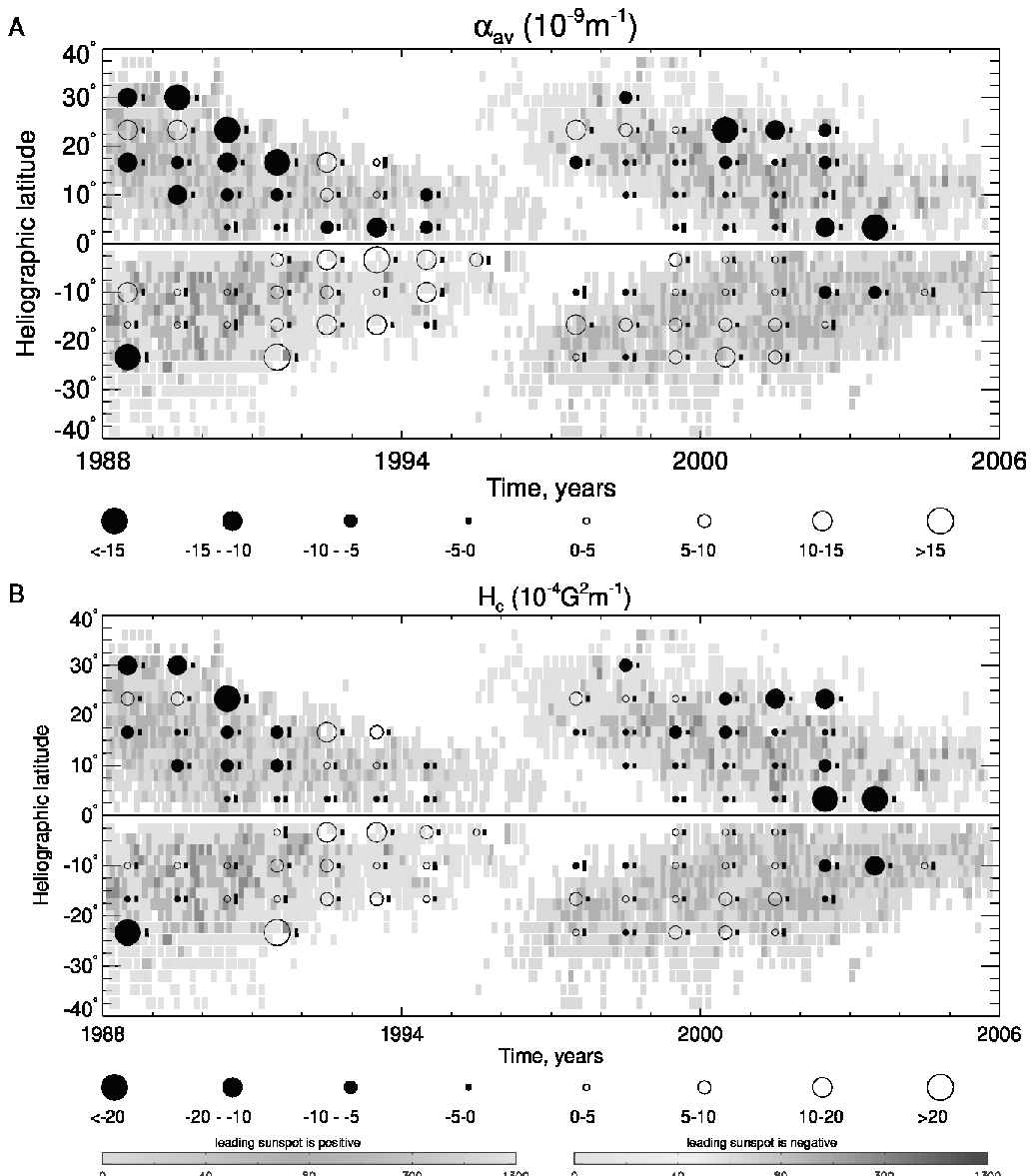}
% }
\caption{ Top: the distribution of the averaged twist $\alpha_{\rm ff}$; and bottom: current helicity $h_{{\rm c}z}$ of solar sunspots in the 22$^{nd}$ and 23$^{rd}$ solar cycles. Superimposed, the underlying grayed ``butterfly diagram'' shows how sunspot density varies with latitude over the solar cycle.   The open (closed) circles indicate the positive (negative) sign.}   
After \cite{zetal10b}\label{fig:butfly1}
\end{figure}

A subsequent study was focused on the distribution of the average helical characteristics of the magnetic field in solar active regions from 1988 to 2005  in the form of butterfly diagrams (latitude-time), which data was observed at Huairou Solar Observing Station, National Astronomical Observatories of China \citep{zetal10b}. It covers most of the solar cycle 22 and 23 in Figure \ref{fig:butfly1}  after removing magneto-optical effects in the measurements of the magnetic field \citep{Su04b,{GZZ08}}. 

The maximum values of the mean helicity density of active regions at the solar surface tend to occur near the edges of the sunspots butterfly diagram. It is roughly consistent with Figure \ref{fig:zb_1998}, where some super active regions are marked. The reversal sign of the mean current helicity relative to the hemispheric sign rule occurs at the beginning of the butterfly wing in 1997 in Figure \ref{fig:butfly1} \citep[also see][]{bao00}. 
It is probably interesting to compare with the traditional solar dynamo theory \citep[such as][]{par55}.

\cite{kuzanyan03,zetal06}  
identified the longitudinal migration of active regions and the corresponding current helicity with their rotation rates and compared with the internal differential rotation law in the solar convection zone inferred from helioseismology. They deduced the distribution of the contributed helicity over depth.

{%\color{red}
 Moreover, \cite{ZhangM06} reported that both $\alpha$ factor and current helicity in the strong fields present a sign opposite to the weak ones, by analyzing large simples of photospheric vector magnetic fields of active regions in 1997 - 2004  observed at the Huairou Solar Observing Station. A similar case with the Spectro-polarimeter (SP) onboard the Hinode satellite was presented by \cite{Hao11}, who showed that the helicity changes the sign from the inner umbra to the outer penumbra.  
The question is how to express the possible reason and mechanism, which relates to the opposite signs of the current helicity between the umbra and penumbra in the same sunspots or the strong and weak fields in the same active regions.
} 
 
%%%%%%%%%%%%%%%%%%%%%%%%%%%%%%
  
Following \cite{MGS82}, it is possible to determine the helicity spectrum of magnetic fields in the solar active regions from the spectral correlation tensor   if we  assume the local statistical isotropy exists. The Fourier transform of the two-point correlation tensor, $\langle B_i({\bf x},t)B_j({\bf x}+{\bf\xi},t)\rangle$ %with respect to ${\bf\xi}$ 
can be written as \citep[cf.][]{M78}
\begin{equation}
\left\langle\hat{B}_i({\bf k},t)\hat{B}_j^*\!({\bf k}',t)\right\rangle
=\Gamma_{ij}({\bf k},t)\delta_2({\bf k}-{\bf k}'),
\end{equation}
where $\hat{B}_i({\bf k},t)=\int B_i({\bf x},t)
\,e^{i{\bf k}\cdot{\bf x}}d^2x$ is the 2D Fourier transform,
the subscript $i$ and $j$  relate to the vector magnetic field components, the asterisk denotes complex conjugation, and ensemble averaging is replaced by averaging over concentric annuli in wavevector space \citep{ZhangBS2016}.  As $\bf k$ defines the direction in $\Gamma_{ij}$, and that $k_i \hat{B}_i=0$, the only possible structure of $\Gamma_{ij}({\bf k},t)$ takes the form  \citep[cf.][]{M78}
\begin{equation}
\Gamma_{ij}({\bf k},t)=\frac{2E_M(k,t)}{4\pi k}(\delta_{ij}-\hat{k}_i\hat{k}_j)
+\frac{iH_M(k,t)}{4\pi k}\varepsilon_{ijk}k_k,\label{eq:helispec5}
\end{equation}
where $\hat{k}_i=k_i/k$ is the component of the unit vector of $\bf{k}$, $k=|\bf{k}|$ is the modulus with $k^2=k_x^2+k_y^2$, and $E_M(k,t)$ and $H_M(k,t)$ are the magnetic energy and magnetic helicity spectra, respectively.

\begin{figure}[!h]
\begin{center}
\includegraphics[width=90mm]{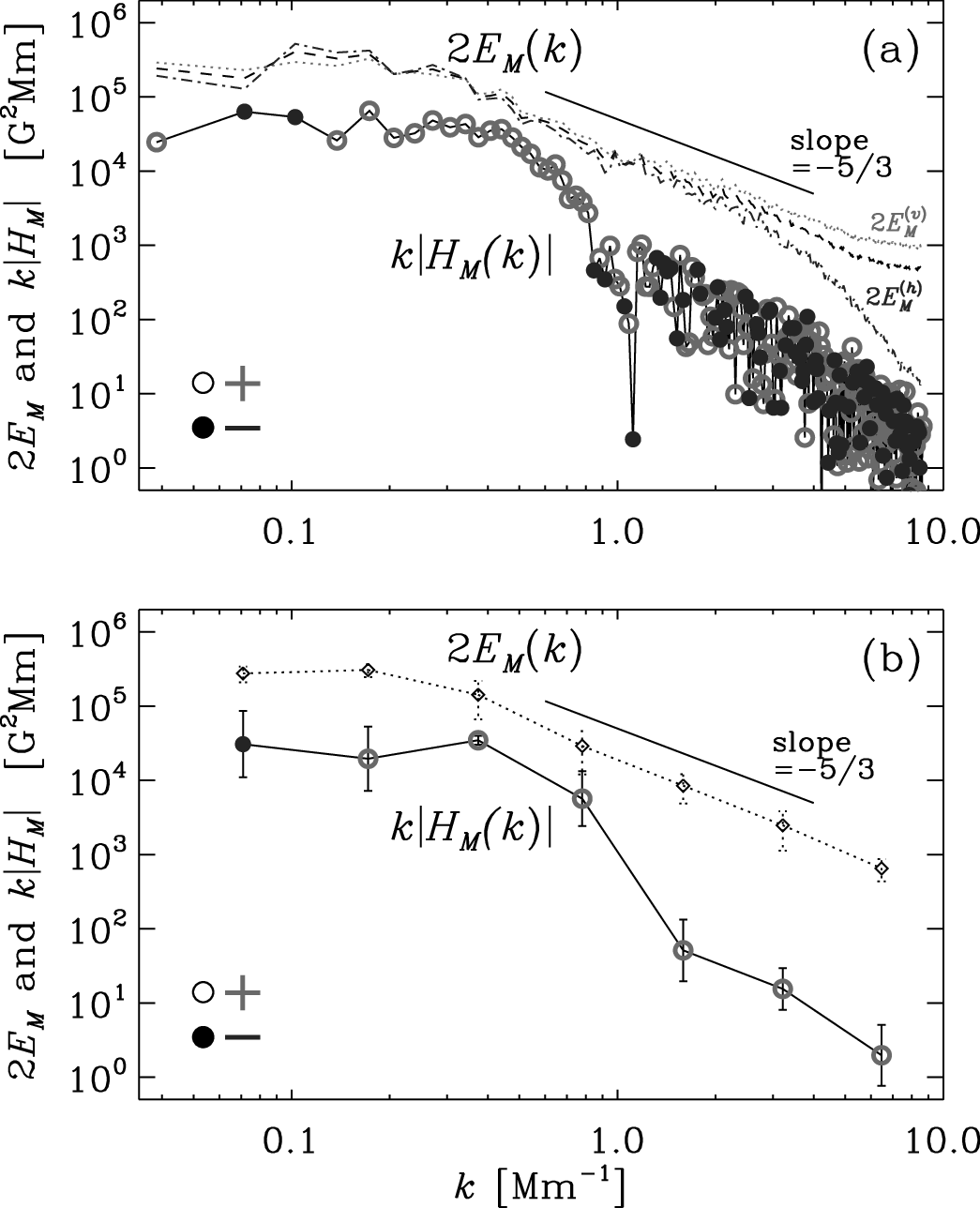}
%\vspace*{1mm}
\end{center}
\caption{(a) $2E_M(k)$ (solid line) and $k|H_M(k)|$ (dotted line) { for NOAA 11158 at 23:59:54UT on
13 February 2011}.
Positive (negative) values of $H_M(k)$ are indicated by open (closed)
symbols, respectively. $2E_M^{(v)}(k)$ (dotted) and $2E_M^{(h)}(k)$ (dash-dotted) are shown for comparison.
(b) Same as the upper panel, but the magnetic helicity is averaged
over broad logarithmically spaced wavenumber bins. After \cite{ZhangBS2014}
\label{fig:phelicity}
}\end{figure}

As an example, Figure~\ref{fig:phelicity} shows the spectrum of magnetic helicity and energy in an active region, which data is observed by HMI onboard the SDO satellite. The lower panel of Figure~\ref{fig:phelicity} shows the sign of magnetic helicity at different scales, after averaging the spectrum overbroad, logarithmically spaced wavenumber bins.

\begin{figure}[!h]
\begin{center}
%\vspace*{-20mm}
\includegraphics[width=90mm]{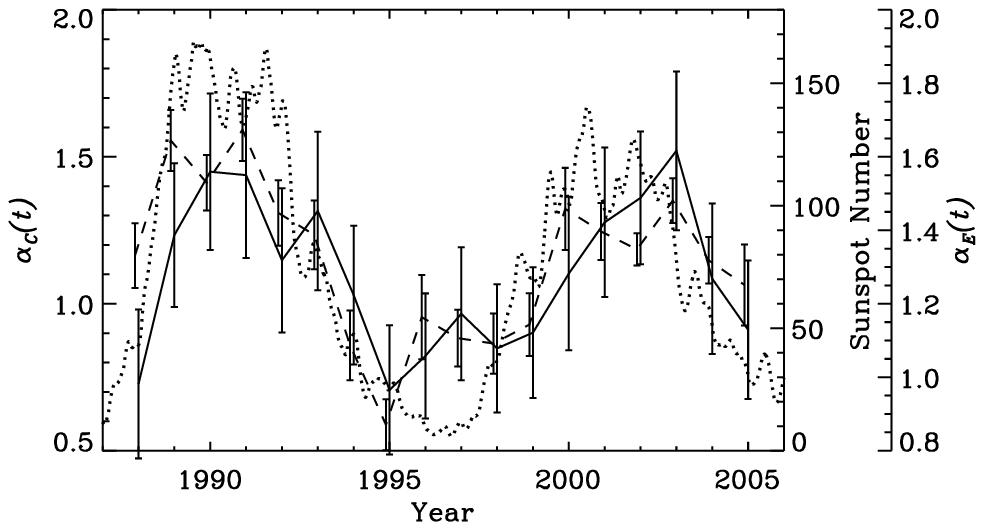}
%\hspace*{-20mm}
\includegraphics[width=80mm]{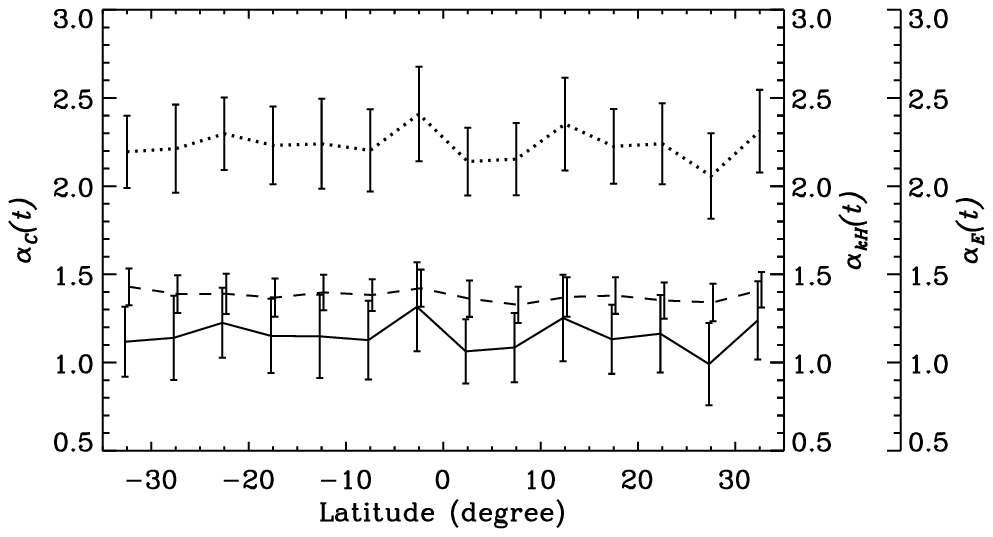}
\caption{ Top: The distribution of average scale exponent $\alpha$  of current helicity ($\alpha_c$ solid line) and magnetic energy ($\alpha_E$ dashed line) with the time inferred by  6629 vector magnetograms of solar active regions in 1988-2005. The dotted line is sunspot numbers.
Bottom: The distribution of average scale exponent $\alpha$  of magnetic helicity ($\alpha_{kH}$ dotted line), current helicity ($\alpha_c$ solid line), and magnetic energy ($\alpha_E$ dashed line) with the latitude. $\alpha_{kH}$ relates to $k|H_M(k)|$.
The error bars  are 0.3$\sigma$. After \cite{ZhangBS2016}}
\label{fig:helispec7}
\end{center}
\end{figure}

Figure \ref{fig:helispec7}  presents the variation of the slopes of the spectrums of magnetic energy and helicities of active regions with solar cycles \citep{ZhangBS2016}.  The magnetic spectrums have been taken in the form of $k^{-\alpha}$. 
The statistical correlation coefficient between sunspot numbers and slopes $\alpha$ of the magnetic energy is 0.827, and that between sunspot numbers and current helicity is 0.730. 
The correlation coefficient between sunspot numbers and slopes of current helicity changes to 0.831, as one takes the sunspot numbers to one year delay. It is consistent with the observed result in Figure \ref{fig:butfly1}, in which the maximum value of mean current helicity of solar sunspots is delayed than that of sunspot numbers. 
Similar evidence is that the complex magnetic configuration of active regions tends to occur in the decaying phase of the solar cycle \citep{Guo10}.

Another is that one cannot find the significant variation of the mean slopes {\color{red}$\alpha$} of the spectrums of magnetic energy and helicities of active regions with the latitudes after a long-term average in Figure \ref{fig:helispec7}. It reflects that the mean scale distribution of the magnetic field of solar active regions doesn't show a significant latitudinal tendency.

A similar study for the active cycle is presented by \cite{Gosain19} after computation of the magnetic helicity and energy spectra from magnetic patches on the solar surface using data observed by the Hinode satellite in 2006-2017.

Similar studies of the large-scale current helicity are based on the calculation of the longitudinal full-disk magnetograms, which are obtained by the Michelson Doppler Imager (MDI) as well as the Kitt Peak Vacuum Telescope and the Synoptic Long-term Investigations of the Sun (SOLIS) by \cite{Pevtsov00,wangcy10,PipinPev14}. The large-scale magnetic fields reflect a clear and consistent current helicity pattern that follows the established hemispheric rule.
\cite{Pipin19} have presented the mean helicity density of the non-axisymmetric magnetic field of the Sun and separated it into the mean large- and small-scale components of magnetic helicity density, which display the hemispheric helicity rule of opposite signs at the beginning of cycle 24. 

\section{Magnetic Helicity Injection in Solar Surface}

\subsection{Magnetic helicity injection inferred from moving magnetic structures}

Magnetic helicity in Eq. (\ref{heli}) is not a directly observable quantity in the solar atmosphere.  The general definition of magnetic helicity does not satisfy the requirement of gauge invariance, and the concept of relative magnetic helicity has been introduced \citep{BergerField84}. The injected magnetic helicity from the solar subatmosphere is a quantity to reflect its relative variation. It can be inferred by the motions of footpoints of the magnetic fields in the solar surface. The change of magnetic helicity   \citep{cha01} 
\begin{equation}
\frac{dH_m}{dt}=-2\oint_{S}({\bf A}_p\cdot {\bf V}_t)B_nds +2\oint_{S}({\bf A}_p\cdot {\bf B}_t)V_nds,\label{eq:heli34}
\end{equation}
where  ${\bf A}_p$ is the vector potential of the potential field, and the subscript $t$ and $n$ mark the transverse and vertical components of magnetic field and velocity field, respectively. The first integral on the right-hand side of eq. (\ref{eq:heli34}) is the contribution from the twisting or shearing motions of footpoints of the magnetic fields on the solar surface, while the second one is that from the emergences of magnetic fluxes from the sub-photosphere \citep{Antiochos87,Kusano02}.

\begin{figure}[!h]
\vspace{-5mm}
\centering
\begin{minipage}[]{0.4\columnwidth}
{\includegraphics[width=1.0\columnwidth]{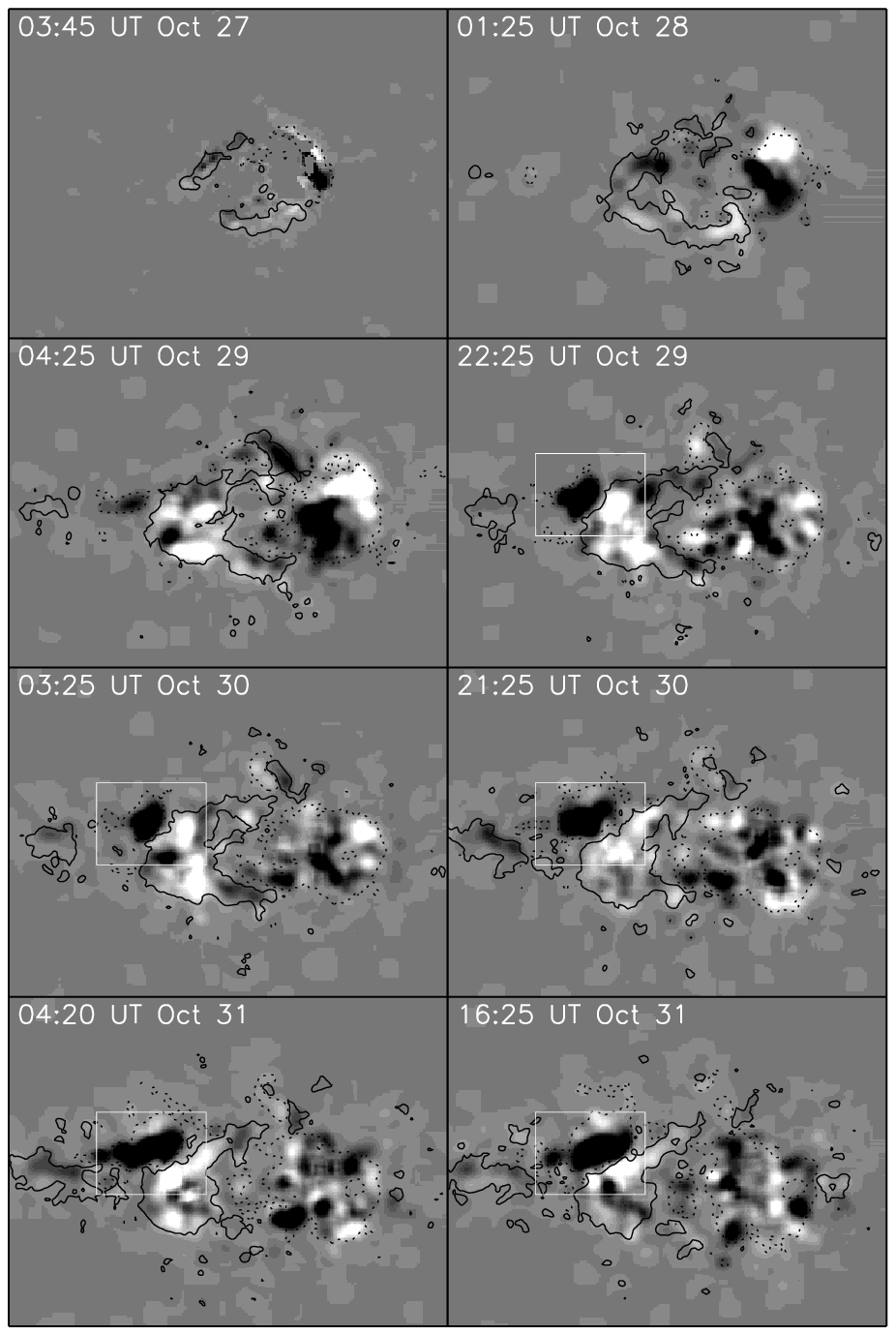}}
  \end{minipage}%
  \begin{minipage}[]{0.54\columnwidth}
\hspace{1.2cm}{\includegraphics[width=0.6\columnwidth]{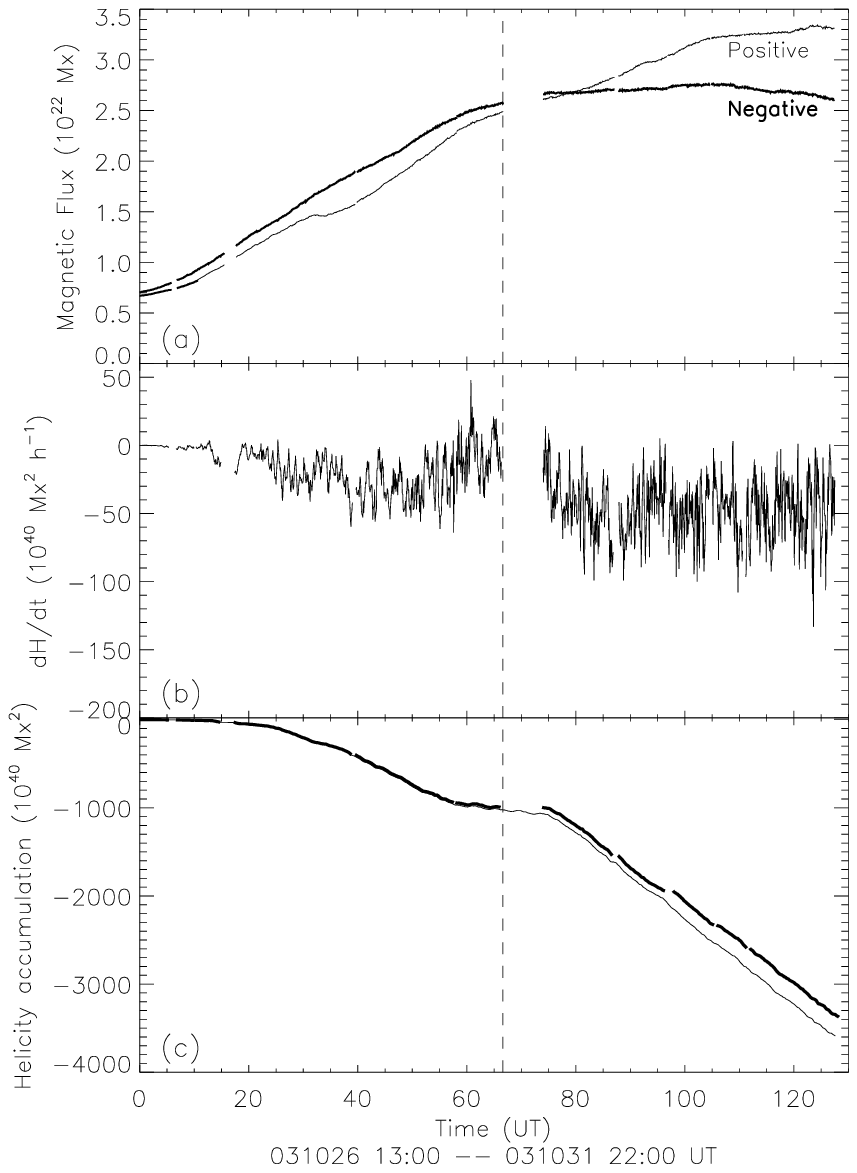}\vspace{0.5cm}}
{\includegraphics[width=0.85\columnwidth]{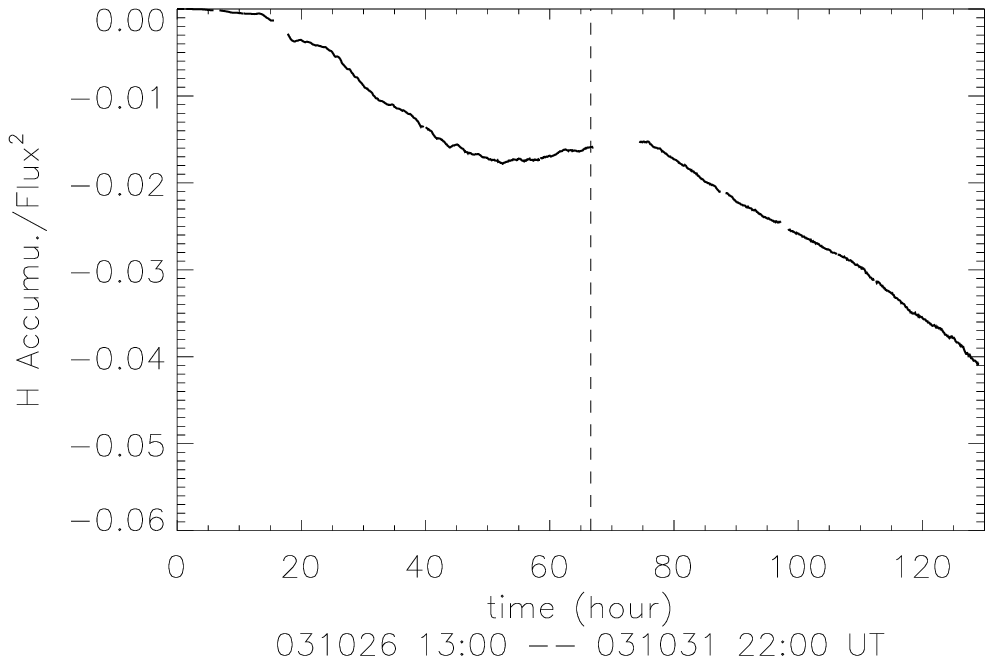}}
  \end{minipage}
  \caption{Active region NOAA 10488 on Oct 26-31, 2003. Left: Gray-scale maps of $G\equiv-2(\mathbf{u} \cdot\mathbf{A}_p)B_z$ of one-hour averages. The white and  black colors show the positive and negative signs, respectively.  Right top: (a) Time profile of the longitudinal component of magnetic field flux. (b) Time profile of the  injected rate of magnetic helicity by horizontal motions. (c) The accumulated helicity $\Delta$H(t) calculated from  ${\mathrm d}H/{\mathrm d}t$.  Right bottom: Ratio of the accumulated coronal helicity to the square of the magnetic flux. After \cite{liu06}
  }
\label{fig:helieme3}
\end{figure}

According to the analysis of \cite{Demoulin03b}, the injected magnetic helicity can be calculated as \citep{cha01}    
\begin{equation}
\frac{dH_m}{dt}=-2\oint_{S}({\bf A}_p \cdot {\bf U})B_nds,
\label{eq:heli35}
\end{equation}
where
\begin{equation}
\label{eq:heli35a}
{\bf U}={\bf V}_t-\frac{V_n}{B_n}{\bf B}_t,
\end{equation}
and ${\bf A}_p$ is magnetic vector potential of the potential field, $t$ and $n$ mark the horizontal and vertical components of the magnetic and velocity field, respectively.
This implies that one can include the most contribution of injective helicity from the horizontal motion of magnetic footpoints in the solar surface.

Figures \ref{fig:helieme3} shows an example of the injection of magnetic helicity in a newly emerging active region by Eq. (\ref{eq:heli35a}).  Both polarities of the magnetic flux increase very quickly in Figure \ref{fig:helieme3}a.  
Figures \ref{fig:helieme3}b and  \ref{fig:helieme3}c display the temporal variation of  ${\mathrm d}H_m/{\mathrm d}t$ and accumulated helicity $\Delta$$H(t)$, separately.  
The former represents the injective rate of magnetic helicity with the horizontal motions of magnetic fields in the solar surface, and the latter represents the total accumulated quantity of helicity.

\subsection{Magnetic helicity injection with solar cycles}

\begin{figure}[htbp]
\begin{center}
\hspace*{-10mm}
\includegraphics[width=120mm]{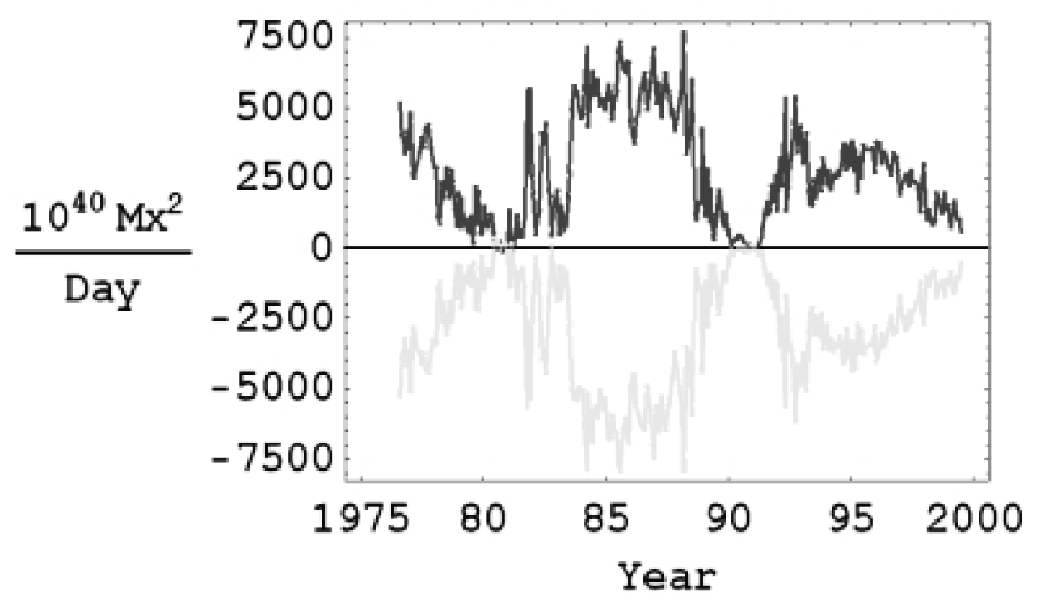}
\end{center}
\caption{Net transfer of helicity into the southern corona and wind $dH_{CS}/dt =
\dot{H}(V_S → C_S)$ (predominantly positive curve), and into the northern corona and wind
$dH_{CN}/dt = \dot{H} (V_N → C_N)$ (predominantly negative curve). The units are $10^{40}
Mx^{2}/day$. From \cite{2000JGR...10510481B} }\label{fig:BR2000-3}
\end{figure}

\cite{2000JGR...10510481B} evaluated the surface integral using solar magnetogram data and differential rotation curves,  
and they proposed the helicity generation in the solar interior by differential rotation to produce the correct sign compared to observations of coronal structures.  
in Figure \ref{fig:BR2000-3}. They estimated that the net helicity flows into each hemisphere over this cycle was approximately $4 \times 10^{46} Mx^{2}$. 
\cite{Georgoulis09} estimated a maximum helicity injection of $6.6\times 10^{45} Mx^2$ for solar cycle 23.  \cite{Hawkes18} applied to data sets covering a total of 60 years to estimate the magnetic helicity as a predictor of the solar cycles,  which is an extension following  \cite{2000JGR...10510481B}.

According to Eq. (\ref{eq:heli35}), the relationship between the mean magnetic helicity density $\overline{h}_m$ and the mean current helicity density $\overline{h}_{cz}$ can be obtained
\begin{equation}
\label{eq:magcurrhelic}
\overline{h}_m%=\frac{1}{S_cL_{c}}\int_{T_{c}}\frac{dH}{dt}dt
=-\frac{2}{S_cL_{c}}\int_{T_{c}}\oint_{\partial S}({\bf U}\cdot {\bf A}_p)B_ndsdt\sim k{\overline{\bf A\cdot B}}\sim L^{2}_c\overline{h}_{cz},
\end{equation}
where $T_c$ is the typical relaxation time,  $S_c$ and $L_c$ are the typical horizontal  and vertical spatial scale of emerging magnetic flux in the solar atmosphere before the transport of helicity into the interplanetary space, and $k$ is a a correlation coefficient (about the order of one).

In principle, it is difficult to obtain $\overline{h}_m$ directly from observation to compare with the current helicity $\overline{h}_{cz}$ in the active regions (in Eq. (\ref{eq:meancurr})). The current helicity $\overline{h}_{cz}$ is inferred from the photospheric vector magnetograms presented above, while Eq. (\ref{eq:heli35}) can be used to      
calculate the injection of magnetic helicity from the subphotosphere. 
The statistical relationship between magnetic and current helicity with solar cycles is probably important to understanding the distribution of magnetic chirality in the solar atmosphere and its transformation from the subatmosphere with solar cycles.

\begin{figure}[!h]
\begin{center}
%\vspace*{-20mm}
\includegraphics[width=120mm]{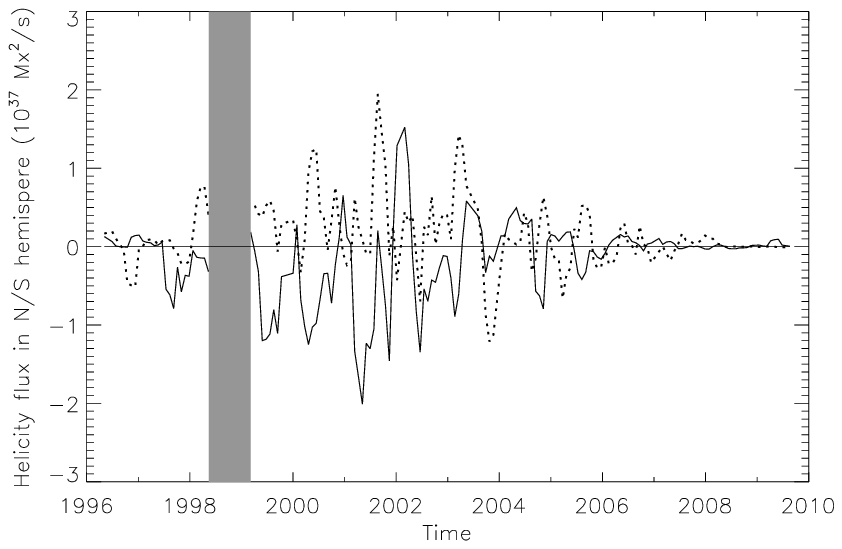}
%\vspace*{-10mm}
%\includegraphics[width=80mm]{hrSza.ps}
\end{center}
\caption{ Injective magnetic helicity flux from the northern (solid line) and the southern (dotted line) hemisphere in 1996-2009.  The shadow areas mark the period without the relevant helicity calculation from the magnetograms. From \cite{Zhang13} }\label{fig:heliflux5}
\end{figure}

For analysis of the injection of global magnetic helicity in the solar surface with the solar cycle, the MDI full-disk magnetograms had been used by  \cite{Yangx12} by tracking the MDI synoptic charts and  \cite{Zhang13} by tracking the full-disk 96 minutes magnetograms.
 The injection of global magnetic helicity in the solar surface can also be calculated or estimated by Eqs. (\ref{eq:heli35}) and (\ref{eq:heli35a}) from the full disk magnetograms as the projective effects have neglected nearby the limb of the solar disk.
Figure \ref{fig:heliflux5} shows the mean injection of magnetic helicity flux in 1996 - 2009 in each solar rotation for analyzing its long-term evolution in the solar surface. 
It can be found that besides the significantly fluctuated injection of magnetic helicity in the northern and southern hemispheres, we also can find the hemispheric sign rule for the injective magnetic helicity \citep{Zhang13}.  It is consistent with the result obtained by photospheric vector magnetograms in the active regions in Figure \ref{fig:butfly1}, due to the negligible contribution from the quiet Sun \citep{Welsch03}. The rough consistency can probably be estimated by Eq. (\ref{eq:magcurrhelic}), even though there may be a complex internal relationship between them.
This may give us enlightenment from the observations, the total injective helicity in a solar cycle tends to be zero, but the fluctuation is significant.

\begin{figure}[!h]
\begin{center}
\includegraphics[width=100mm]{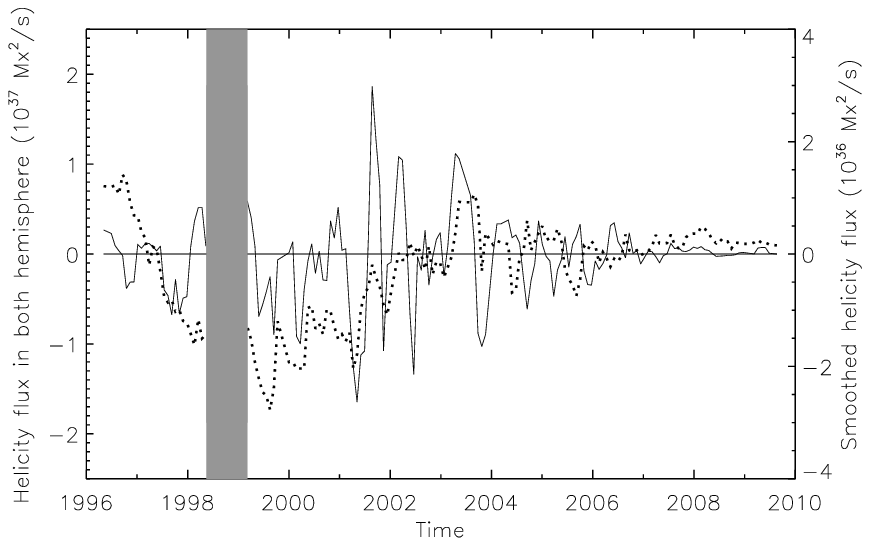}
\vspace*{-10mm}
\includegraphics[width=100mm]{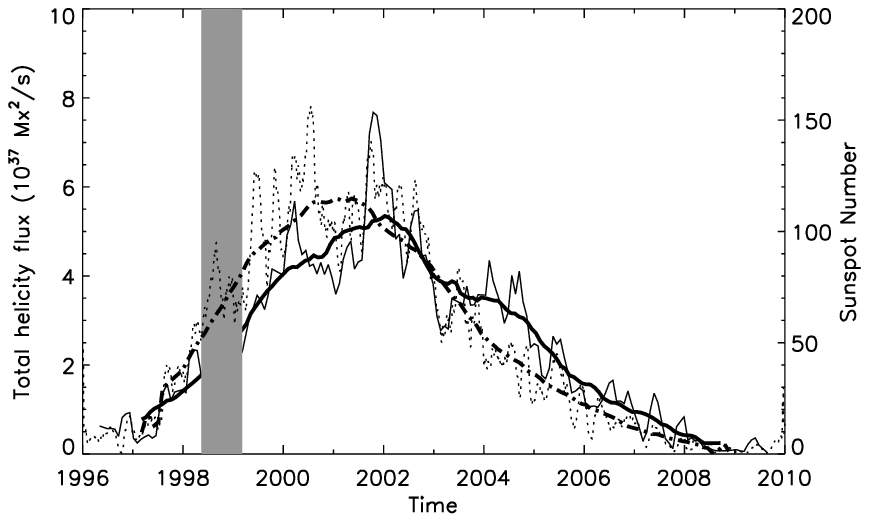}
\end{center}
\caption{Top: The net injective helicity contributed from both hemispheres. The dotted line marks the net injective helicity after the smooth of 48 solar rotations. Bottom: Total magnetic helicity flux (thin solid line) and sunspot numbers (thin dotted line) in 1996-2009, while the black thick solid (dot-dashed) line shows the total helicity flux (sunspot numbers) after smoothed significantly.
From \cite{Zhang13}
}\label{fig:heliflux6}
\end{figure}

It is found that the extreme value of negative helicity flux occur in 1997-2002, after the smooth of net magnetic helicity flux in Figure \ref{fig:heliflux6}. 
 It provides a rough estimation only, the extreme value is about $-2\times 10^{36}Mx^2s^{-1}$.  
It also depicts the injective rate of the total absolute value of magnetic helicity flux inferred from the solar surface and the sunspot numbers in 1996-2009.  The injective rate of the total absolute value of helicity flux is obtained from the absolute value of positive and negative ones.  The mean value of the absolute injective rate of total helicity is $2.40\times 10^{37}Mx^2s^{-1}$ in the calculated solar disk in 1996-2009.
This provides a basic estimation of total injective magnetic helicity flux in about $5.0\times 10^{46}Mx^2$ in the 23rd solar cycle from both hemispheres, and it is a similar order as estimated by \cite{2000JGR...10510481B} and also  \cite{YangSB12}.
It may provide a minimum value if the contribution of small-scale helicity flux and the projective effects of the magnetic field in the solar surface has been estimated in the solar cycle 23.

 It is also noticed that the total magnetic helicity flux tends to delay than the total sunspot numbers, as comparing both smoothed ones in Figure \ref{fig:heliflux6}. The maximum of sunspot numbers occurs in 2001, while that of current helicity in 2002 after the smooth. This is consistent with that the maximum of butterfly diagram of calculated current helicity of solar sunspots delays than that of sunspot numbers in Figure \ref{fig:butfly1}.
It also means that the sunspot numbers do not reflect the relevant handedness of magnetic fields generated in the solar atmosphere completely.

%%%%%%%%%%%%%%%%%%%%%%%%%%%%%%%%%%%%%%%%%%%%%%

\section{Magnetic Helicity as an Index in the High Solar Atmosphere}

When we study the upward transport of magnetic helicity from the solar photosphere, the helical characteristics of the coronal magnetic field can give us important instructions. Under the existing observational conditions, the chirality study of the coronal loops is of great significance.

\subsection{Hemispheric Distribution of Helical Coronal Soft X-Ray Loops}

\begin{figure}[h]
\begin{center}
\vspace{-10mm}
\includegraphics[width=100mm]{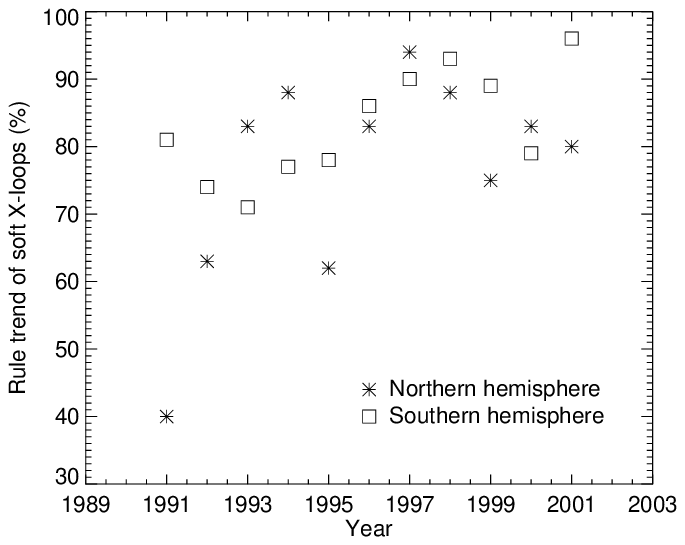}
\end{center}
\caption{The hemispheric handed rule trend of the proportion of soft X-ray loops  in the southern and northern hemispheres. From \cite{Zhangetal10a} }\label{fig:synARb6}
\end{figure}

To analyze the magnetic chirality in the solar corona, Figure \ref{fig:synARb6} shows the statistical results of 753 large-scale soft X-ray loops in 1991 - 2001 observed by the Yohkoh satellite. 
As the unidentified loops are ignored, one can find that the portion of the systems which are in accord with the hemispheric rule is $77.3\%$ ($81.5\%$) in the northern (southern) hemisphere. It is roughly consistent with the hemispheric sign rule of the magnetic helicity in active regions.

%\begin{table}
% \caption{Handedness statistics of solar corona soft X-ray loops in the southern (northern) hemispheres\label{tab:synARb1} }{%
%\begin{tabular}{@{}ccccccccccccc@{}}
%\toprule
%\toprule
% Year  &1991&1992&1993&1994&1995&1996&1997&1998&1999&2000&2001&Total\\
%\toprule
%$N n$ & 4 & 38 & 25 & 31 & 5 &5 & 16&23 &32 &24 &22 &225\\
%$P n$ & 6 & 24 & 5  & 4 &3 & 1& 2& 3& 9&6 &5 &68 \\
%$Q n$ & 7 & 27 & 22 & 14& 5& 5& 11& 15&18 &13 & 7 &144\\
%\toprule
%%\tableline
%$P s$ & 13 & 23 & 24 & 13 & 7 & 6 & 19 & 26 & 16 & 11 & 27&185 \\
%$N s$ & 7 & 8 & 10 & 4 & 2 & 1 & 2 & 2 & 2 & 3 & 1 &42 \\
%$Q s$ & 3 & 12 & 4 & 5 & 7 & 4 & 5 & 22 & 14 & 6 & 7&89 \\
%\toprule
% $Total$&40 & 132 & 90 &  71& 29&22 & 55 & 91 & 91 & 63 & 69 &753\\
%\toprule
%\end{tabular}}
%\footnotetext{$P$ is the number of soft X-ray loops with right-handedness, $N$ is the number of soft X-ray loops with left-handedness, $Q$ is the number of unidentified soft X-ray loops.\\
%The subscript $n$ and $s$ indicate the northern and southern hemispheres, respectively.}
%\end{table}

\begin{figure}[!h]
\begin{center}
\vspace{-10mm}
\includegraphics[width=90mm,angle=0]{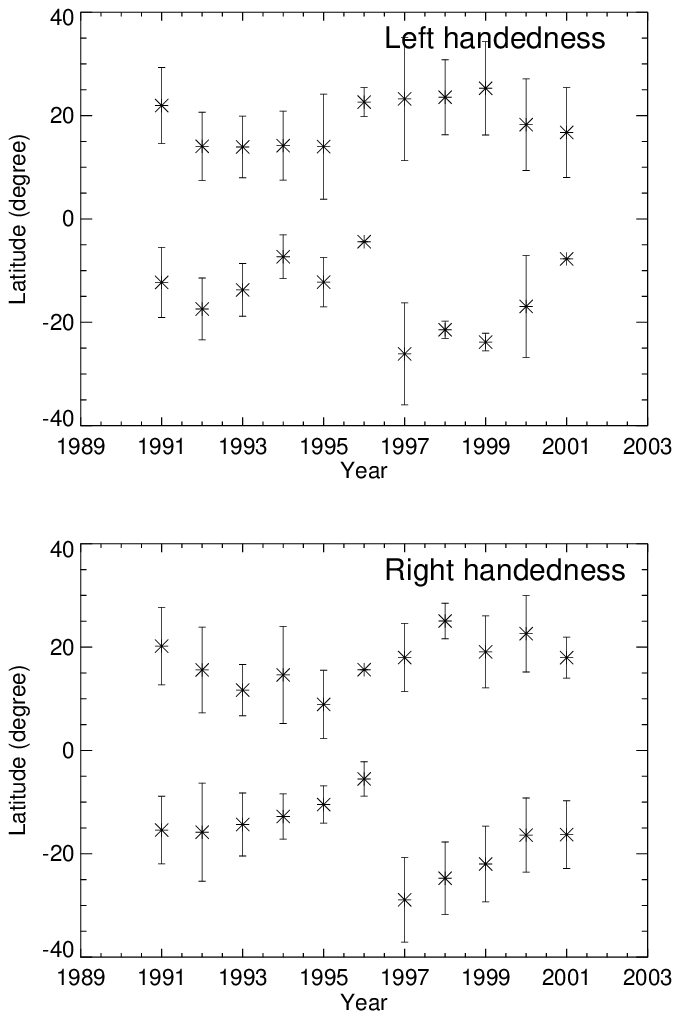}
\end{center}
\vspace{-10mm}
\caption{The mean latitudinal distribution of soft X-ray loops with left and right-handedness. $\sigma$-error bars are shown by vertical lines. From \cite{Zhangetal10a} }\label{fig:synARb7}
\end{figure}

Figure \ref{fig:synARb6} also depicts %the proportion of soft X-ray loops following the hemispheric handedness rule in the northern (southern) hemisphere. It is found that the change in the proportion of soft X-ray loops following the hemispheric handedness rule of helicity and also
the imbalance of chiral soft X-ray loops in both hemispheres. The reverse magnetic helicity with a relatively high tendency occurred in 1991, 1992, and 1995 in the northern hemisphere, while it is insignificant in the southern hemisphere.

Figure \ref{fig:synARb7} shows the statistical distribution of the helical soft X-ray loops with the latitudes. The mean latitude of soft X-ray loops migrates toward the equator with the solar cycle in the form of a butterfly diagram. It is similar to the distributed form of the sunspots. Due to very few soft X-ray loops in 1991, 1995, and 1996 in our statistics, the deviation from the butterfly diagram in these years can be noted.  Besides most of the large-scale loops following the helicity hemispheric sign rule, the statistical distribution of the reverse helical soft X-ray loops relative to the sign rule can be found in Figure \ref{fig:synARb7} also.

\begin{figure}[!h]
\includegraphics[width=60mm]{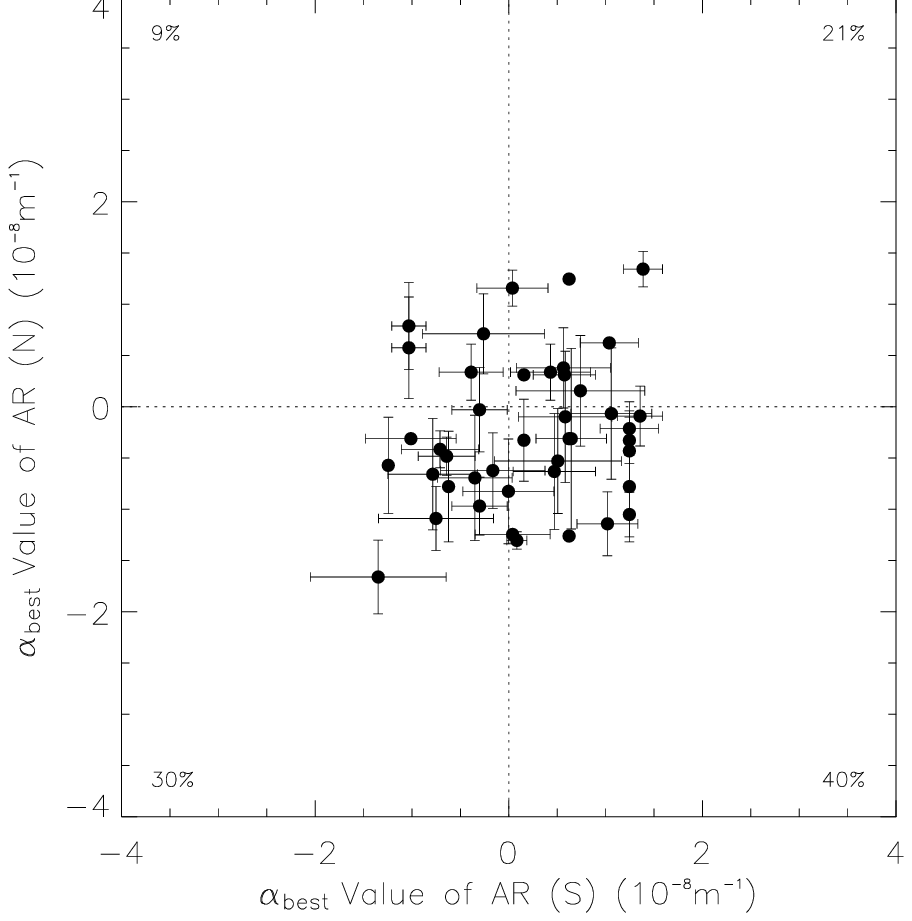}
\includegraphics[width=60mm]{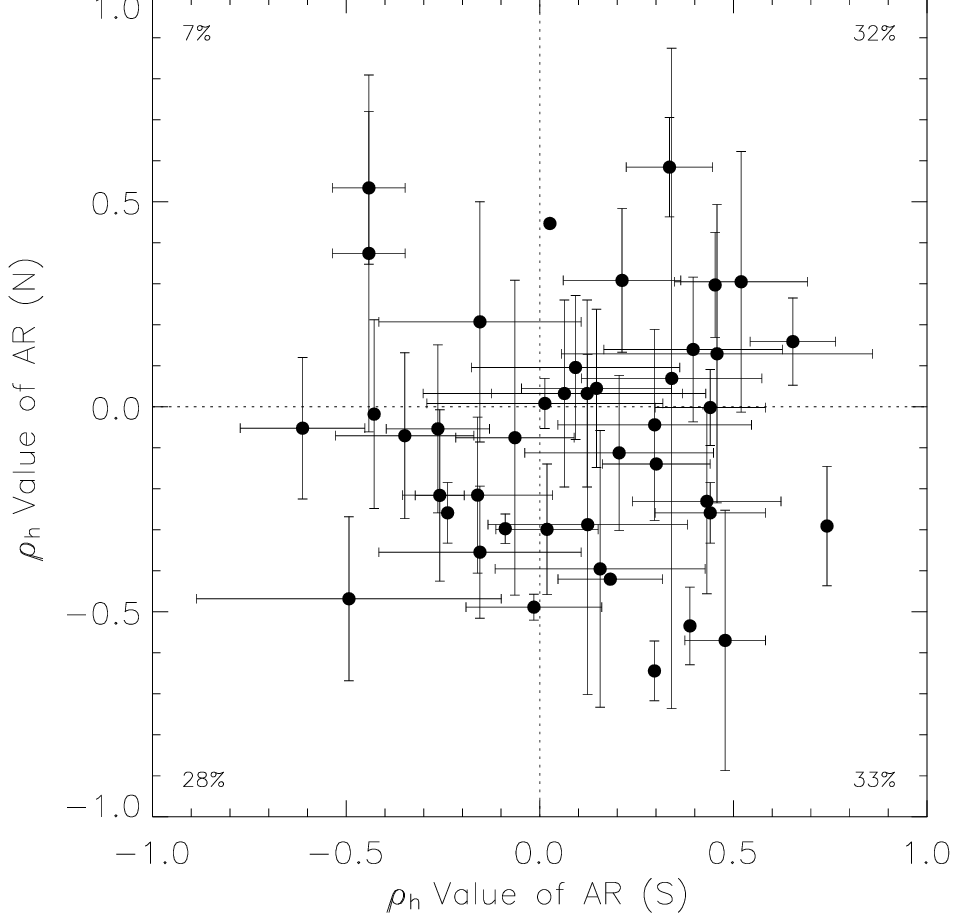}
\caption{ {\it Left:} Correlation of $\alpha_{best}$  of the active region pairs connected by transequatorial loops. {\it Right:} Relationship of $\rho_h$ of the active region pairs connected by TLs. X axis presents the helicity values of active regions in the southern hemisphere, Y axis shows the values in the northern hemisphere. Error bars (when present) correspond to $1\sigma$ of the mean helicity values from multiple magnetograms of the same active
region. Points without error bars correspond to active regions represented by a single magnetogram. After \cite{Chen07} }\label{fig:tranloopb3} 
\end{figure}
 
Moreover, from the time scale of solar cycles, the trans-equatorial X-ray loops connected the solar active regions from both hemispheres and the relevant current helicity of these pair regions have been statistically provided by \cite{Chen07}. 
Figure  \ref{fig:tranloopb3} exhibits the helicity correlation of the 43 pairs of active regions which are connected by transequatorial loops \citep{Chen07}. %From the figure, we can't see the obvious relationship between them. 
For the helicity parameter $\alpha_{best}$ ($\sim\overline{(\nabla\times {\bf B})_{z}/B_{z}}$ in Eq. (\ref{eq:meanalpha})), 22 pairs (51\%) of active regions show the same helicity patterns and 21 pairs (49\%) of active regions own the opposite chirality. For the parameter $\rho_h$ ($\displaystyle ={\sum{h_c(i,j)}}/{\sum{|h_c(i,j)|}}$),  26 pairs (60\%) of active regions show the same signs and 17 pairs (40\%)  the opposite signs. The results of both proxies exhibit that the active region pairs connected by transequatorial loops do not necessarily contain the same chirality.
It reflects the large-scale poloidal magnetic field in the high solar atmosphere. Whether these loops contain more information about the global transformation of magnetic helicity with solar cycles is worthy of further study.

\subsection{Helicity with  Solar Flare Cycles}

The relationship between magnetic helicity and the activity of the flare-coronal mass ejections is a notable topic. It relates to the transformation of magnetic helicity from the solar atmosphere into the interplanetary space \citep[cf.][]{Rust02,Yangx12,wangcy15,Kim17,Park21}.
%In the panel of $|\alpha_{av}|$ of Figure~\ref{F-meansdev}, the values mostly concentrate in the range of $0\thicksim 1.5 \times10^{-5}$ km$^{-1}$. There is also no distinct difference in the parameter $|\alpha_{av}|$ during the evolution proceeding of solar cycles. The difference of $|\alpha_{av}|$ between flare-productive samples and flare-quiet samples is insignificant except that in the years 2004 and 2005.

\begin{figure}[!h]
\begin{center}
              \includegraphics[width=0.7\textwidth,clip=]{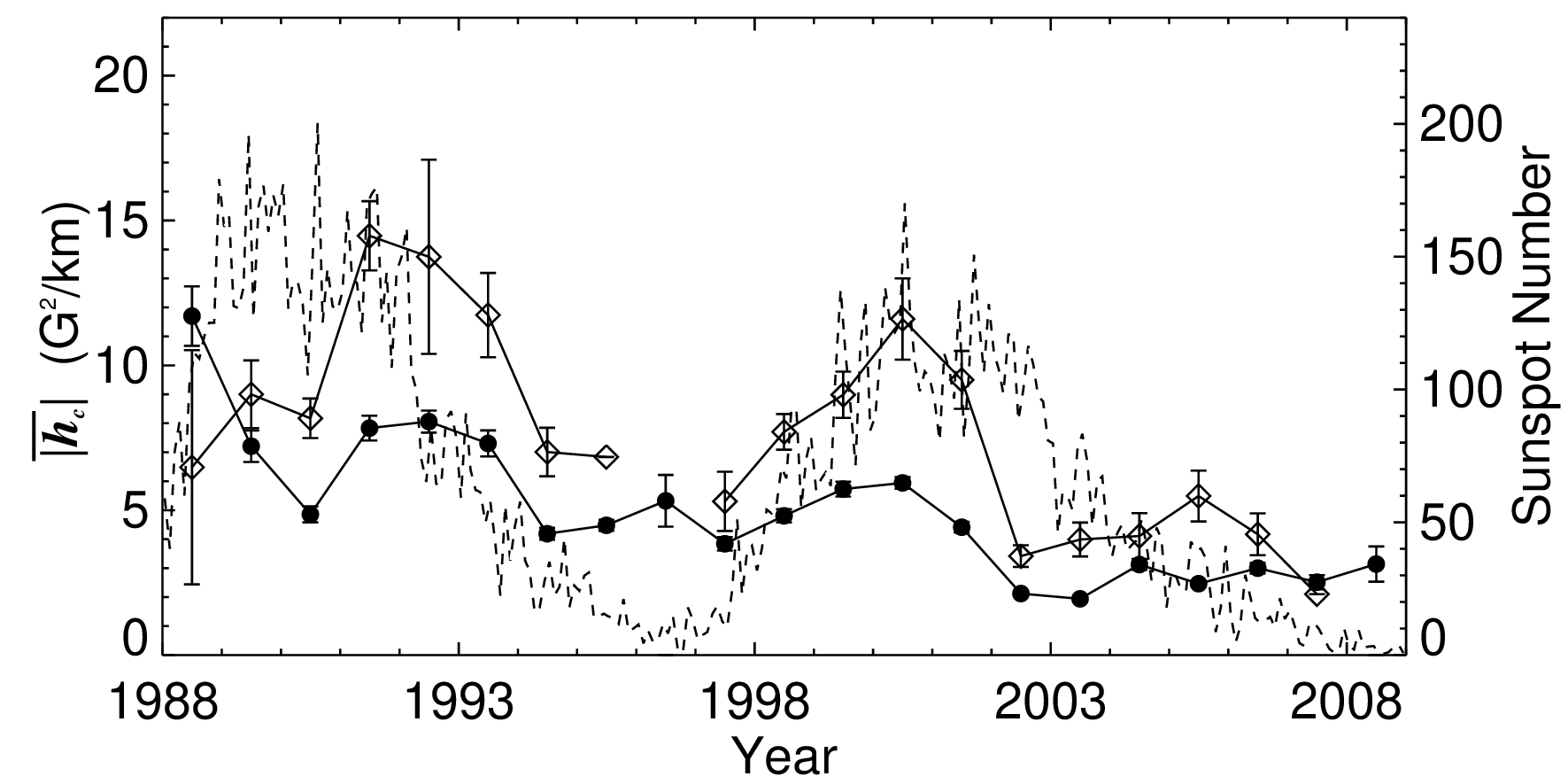}
              \includegraphics[width=0.7\textwidth,clip=]{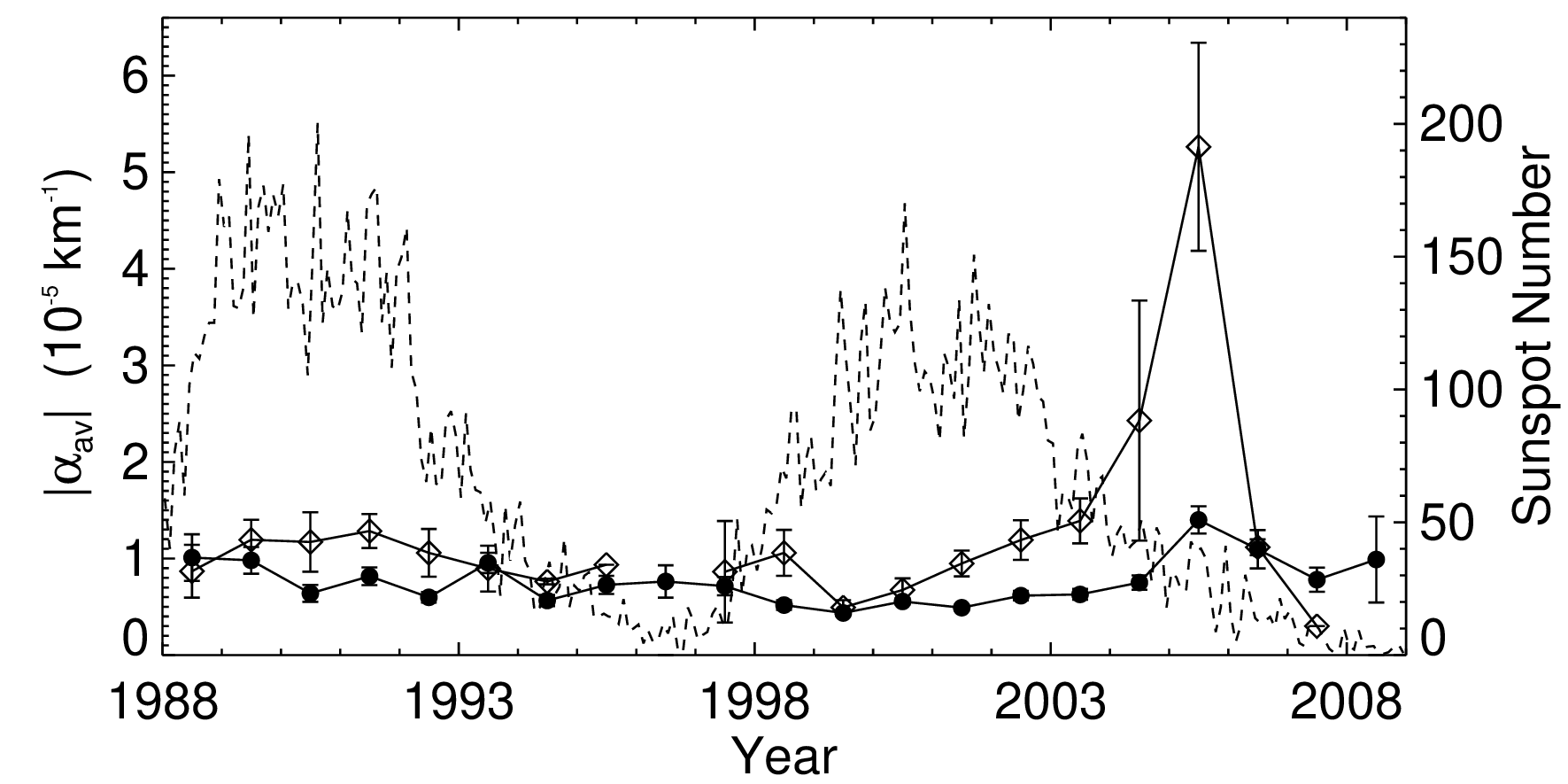}
\end{center}
\caption{Yearly mean values of $\overline{|h_{cz}|}$ and $|\alpha_{av}|$
of AR samples during 1988-2008.  Diamonds represents the yearly mean values of the samples that produced flares with $\textrm{FI}\geq 10.0$ in the following 24 hours (flare-productive samples). Dots represents the yearly mean values of the samples that did not produce flares with $\textrm{FI}\geq 10.0$ in the following 24 hours (flare-quiet samples). The monthly mean sunspot relative numbers during the same period are also overlapped (dashed line).  From \cite{Yangx12}  }
     \label{F-meansdev}
\end{figure}

%The parameter $\overline{|h_{c}|}$ follows the solar cycle very well. There are two bell-shaped parts in its panel during these two solar cycles. The overall level of $\overline{|h_{c}|}$ raises gradually towards the solar maximum and it falls near the solar minimum. This result is also consistent with \cite{BZ98} who have surveyed the evolution of the average current helicity in solar cycle 22.

%From 1975, the GOES satellite has been recording the whole-sun X-ray fluxes at 0.5-4 {\AA} (hard channel) and 1-8 {\AA} (soft channel) wavelength bands. The SXR flare classification (B, C, M, and X classes) utilizes the flux in the 1-8 {\AA} range based on the order of magnitude of the peak burst intensity. During a fixed time window $\tau$, the SXR flare index by summed  of flares classes is usually used to evaluate the energy release of an AR:
%\begin{eqnarray}
%\textrm{FI}=100\sum_{\tau} I_{X}+10\sum_{\tau} I_{M}+\sum_{\tau} I_{C},
%\end{eqnarray}
%where $I_{X}$, $I_{M}$, and $I_{C}$ represent the indexes of X-, M-, and C-class, respectively (\citealp{Antalova96,Ab05}). Here we negelect the records of B-class SXR flares, becasue the background in the solar maximum is too high to detect B-class flares (cf., \citealp{Feldman97}.

For statistical studying the photospheric magnetic nonpotentiality in solar active regions and its relationship with associated flares, \cite{Yangx12} selected 2173 photospheric vector magnetograms from 1106 active regions (ARs) observed by the Solar Magnetic Field Telescope at Huairou Solar Observing Station in 1988 – 2008, which covers most of the 22nd and 23rd solar cycles.
A series of parameters have been calculated, which include the mean absolute current helicity density, absolute averaged twist force-free parameter, mean free magnetic energy, and other non-potential parameters  
of each vector magnetogram.

\cite{Yangx12} selected the active samples (i.e., flare-productive ARs) are defined as the ARs with the equivalent flare strength greater than a typical value (such as M1.0 flare here) within the same subsequent time window (24 hours). The rest of them belong to the quiet samples (i.e., flare-quiet ARs). 

In Figure~\ref{F-meansdev}, the mean current helicity $\overline{|h_{cz}|}$ of the flare-productive active regions shows statistically higher values than those for flare-quiet ones in the solar maximum. However, the twist factor $|\alpha_{av}|$ of active regions shows an insignificant difference between the quiet and the active samples, in comparison with $\overline{|h_{cz}|}$. However, it is noticed that the mean helicity density $\overline{|h_{cz}|}$ of the flare-productive active regions shows two delayed peaks relative to the sunspot maximum in 1992 and 2005. A similar case of $|\alpha_{av}|$ can also be found in 2005, respectively. It is consistent with the results calculated from the magnetic helicity in Figures \ref{fig:helispec7} and \ref{fig:heliflux6}. It probably reflects that the formation on the peaks of the flare-productive active regions with strong magnetic helicity tends to be statistically delayed than that of sunspot numbers.

\section{ Questions on Solar Dynamos with Observation of Magnetic Helicity}

It is noticed that exploring the variation of magnetic helicity with the solar cycles from different perspectives is of great significance to understanding the law of formation of the solar internal magnetic field, which relate to the emergence of twisted magnetic flux in active regions \citep[c.f.][]{lon98,Fisher00,ketal20,kuzanyan20}.

In comparing with observations of magnetic helicity, \citet{Choudhuri03} argued that an extremely important question is whether, from a solar dynamo perspective, is it merely a statistical fluctuation of current helicity, or is there some systematic aspect in it, whether certain current helicity is preferential at certain latitudes and time appears.
 Our paper's more extensive data analysis provides a chance to settle this question. On theoretical grounds, they expected that the flux tubes may have corresponding current helicity systematically in certain latitudes at certain times. They suggested that flux tubes at the bottom of the convection zone are advected equatorward by the equatorward meridional flow there, whereas the poloidal field at the surface is advected poleward by the poleward meridional flow there. A systematic study of the observed signs of current helicity in different latitudes in different phases of the solar cycle should throw important light on the nature of the solar dynamo. The extended numerical simulation for the solar dynamo with helicity has been presented by \citet{Choudhuri04}.   

A similar presentation is that   \citet{Xu09} employed a very simple Parker dynamo model and direct generalizations of the latitude time distribution of the helicity obey a polarity rule that is similar to Hales polarity rule. There is a stable phase shift between the helicity and the toroidal magnetic field in the dynamo model. Their dynamo models are more sophisticated than Parker's simplest model which could predict phase shifts and butterfly diagrams. It is natural to suppose that a description of the evolution of the helicity in terms of a differential equation could considerably modify the phase shift between the helicity and the toroidal magnetic field to compare with the observed variations in the magnetic field and helicity with solar cycles \citep{bao00,zetal10b}.

It is defined by the evolution of the averaged small-scale magnetic helicity, $\overline{\chi}=\overline{\mathbf{a}\cdot\mathbf{b}}$ ($\bf{a}$ and is $\bf{b}$ fluctuating part of the magnetic vector potential and magnetic field respectively). For the isotropic turbulence \citep{M78}, the current helicity is related with  magnetic helicity, $h_{c}={\overline{{\bf b}\cdot{\nabla\times  {\bf b}}}\sim\overline{\chi}}/\ell^{2}$.
 The evolution equation for $\overline{\chi}$ can be obtained from the equations which govern the evolution of  $\mathbf{a}$ and $\mathbf{b}$, it reads as follows \citep{kle-rog99}:
\begin{eqnarray}
\label{eq:hel} 
\frac{\partial\overline{\chi}}{\partial t} & = & -2\left(\boldsymbol{\mathcal{E}}\cdot\overline{\bf{B}}\right)-\frac{\overline{\chi}}{R_{m}\tau_{c}}-\boldsymbol{\nabla}\cdot\boldsymbol{\boldsymbol{\mathcal{F}}}-\eta\overline{\mathbf{B}}\cdot\mathbf{\overline{J}},
\end{eqnarray}
where        
 $\boldsymbol{\mathcal{E}}=\alpha_{0}\overline{\mathbf{B}}+\left(\mathbf{\overline{V}}^{(p)}\times\overline{\mathbf{B}}\right)-\eta_{T}\left(\nabla\times\overline{\mathbf{B}}\right)$, and $\overline{\mathbf{B}}$, $\overline{\mathbf{J}}$ and $\overline{\bf {V}}^{(p)}$ are mean part of the magnetic field, current and velocity field respectively, the equipartition value ${\displaystyle \overline{B}_{\rm eq}\sim\sqrt{4\pi\overline{\rho}\overline{u^{2}}}}$, $R_{m}$ is magnetic Reynolds number, and  the helicity fluxes $\boldsymbol{\boldsymbol{\mathcal{F}}}=\mathbf{\overline{a\times u}}\times\mathbf{B}-\mathbf{\overline{a\times(u\times b)}}- \overline{\mathbf{b}\phi}$,  $\phi$ is an arbitrary scalar function which is related with a gauge of the vector potential  (see \citealp{Kleeorinetal95,kle-rog99} and references therein).
The subsurface kinetic helicity in solar active regions has been inferred from the helioseismology \citep[e.g.][]{GZZ09,GZZ12,Komm19}, which tends with the opposite sign relative to the current helicity statistically \citep[e.g.][]{GZZ09,GZZ12} (to the hemispheric current helicity sign rule).

The relatively small time scale fluctuation or oscillation of mean magnetic helicity in the solar cycles has been performed in  Figures \ref{fig:butfly1},  \ref{fig:heliflux5},  and  \ref{fig:heliflux6}.  Some of  the theoretical explanations for the evolution of magnetic helicity in eq. (\ref{eq:hel})  in the framework of mean-field dynamo can be found \citep[][]{ketal03,zetal06,2012ApJ...751...47Z,YangS20} in comparing with the observations.
These reflect that the generation of the magnetic field is a relatively complex process in the convection zone, and one probably cannot use a simple model to explain the twisting process of the magnetic field inside of the Sun. It also provides us an important opportunity to study the solar cycle variation of the magnetic helicity from the perspective of the characteristics of magnetic turbulence based on the solar magnetic fields \citep[][]{Hoyng93,Zhang12}.

The importance of the magnetic helicity with the kinetic helicity and the solar dynamo has been noticed \citep[cf.][]{pouquet-al:1975,kleruz82,bra-sub:05,rad07}.  
The $\alpha$ effect is produced by the kinetic helicity and also current helicity, and it is ${\displaystyle \alpha_{0}=-\frac{\tau_{c}}{3}\left(\overline{{\mathbf u}\cdot{\nabla\times}{\mathbf u}}-\frac{\overline{{\bf b}\cdot {\nabla\times {\bf b}}}}{4\pi\overline{\rho}}\right)}$. The latter effect is interpreted as a resistance of magnetic fields against  a twist by  helical motions \citep{Vainshtein92,rue93}. This effect has been introduced a concept as called ``catastrophic quenching'' of the $\alpha$ effect related to the generated large-scale magnetic field. It was found that        
 $${\displaystyle
   \alpha_{0}\left(\overline{B}\right)=\frac{\alpha_{0}\left(0\right)+\eta
   R_m\overline{\bf B}\cdot \overline{\mathbf J}}{1+R_{m}\left(\overline{B}/\overline{B}_{\rm eq}\right)^{2}}}.$$
 In the case of $R_{m}\gg1$, the $\alpha$ effect
is quickly saturated for the large-scale magnetic field if the strength is much below the equipartition value $\displaystyle \overline{B}_{\rm eq}$ \citep{oss2001}.  

Furthermore, we may not simply relate many questions in the solar dynamo mechanism \citep[c.f.][]{Moffatt19} from the observational perspective of magnetic helicity, but a statistically observed magnetic helicity with the solar cycle provides constraints for this study. A further question is to what extent we can really diagnose the mechanism inside the sun through the observed transmission and evolution of magnetic helicity in the solar atmosphere.

\section{Discussions}

% The helical topology of magnetic fields in active regions was firstly observed from the handedness of sunspot penumbral configuration by \cite{Hale08} and statistically with the hemispheres by \cite{Ding87}.  The hemispheric helicity sign rule in the solar atmosphere was presented \citep[e.g.]{Seehafer90} and also statistically relate to the subsurface kinetic helicity in solar active regions with the opposite sign's tendency  \citep[e.g.][]{GZZ09,Komm19}.

In this chapter,  we have presented some relationships between magnetic helicity and solar cycles.  We have introduced the magnetic helicity and current helicity and their possible relationship from the solar observations. %We have presented the calculation of magnetic and current helicity inferred from the observed magnetic field. 
For analyzing the relationship between the magnetic helicity with solar cycles, we introduce the butterfly diagram of the mean current helicity density inferred from the vector magnetic fields of active regions %observed at the Huairou Solar Observing Station 
and the relationship with the sunspot butterfly diagram in Figure \ref{fig:zb_1998} and \ref{fig:butfly1}, and soft X-ray loops in Figure \ref{fig:synARb7}. 
Another notable topic is the relationship between the magnetic (current) helicity and solar flare-coronal mass ejections with solar cycles. It has presented in Figure \ref{F-meansdev}. We also have presented the injection of magnetic helicity with solar cycles inferred from full-disk magnetograms from the space and ground bases, which provides an aspect of the magnetic helicity that differs from the current helicity calculated by the photospheric vector magnetic field in solar active regions. 
The message which we infer from the magnetic helicity with solar cycles is as follows:

\begin{enumerate}
 \item The twist and helicity butterfly diagrams of active regions show anti-symmetric relative to the solar equator. It confirms the hemispheric rule of magnetic helicity of active regions in 11-year cycles \citep[such as][]{Ding87,Pevtsov95,BZ98,ketal03,Zhang13,liu22}.
 It actually reveals the  new veil on the dynamo process of the magnetic field inside of the Sun.

\item The study of current helicity in the solar surface in eq. (\ref{eq:meancurr}) and injective magnetic helicity through the solar surface in eq. (\ref{eq:heli35}) reflect different aspects of the twisted magnetic field with handedness. The basic relationship between them can also be estimated in the simple form in eq. (\ref{eq:magcurrhelic}), so one can find a similar tendency in that the magnetic and current helicity also shows a temporal delay of the maximum of helicities relative to that of sunspot number in the solar cycles in Figures \ref{fig:helispec7} and \ref{fig:heliflux6}. 

 \item The coronal soft X-ray loops reflect the extension of the photospheric magnetic field. It is found that the handedness of most of the loops tends to follow the hemispheric rule. The interesting question is how the twisted magnetic field extends from the photosphere into the corona within the solar cycles. A similar question is also for the quiescent and active filaments in the solar atmosphere \citep{Martin94,Ouyang17}, 
and also how the magnetic helicity spreads out from the active region, and whether there are other possible causes.

 \item The variation of magnetic helicity with solar flare activity was reported by \cite{Bao99}. We have presented some results on the statistical correlation between them with solar cycles. However, the complexity between them is not completely clear, because more magnetic parameters need to be optimally analyzed for the solar flare-eruptive phenomena \citep{Yangx12}. It is also noticed that the statistical magnetic (current) helicity related to flares also shows a temporal delay than sunspot number with solar cycles in Figure \ref{F-meansdev}. 
 Moreover, the relationship between solar flare-coronal mass ejections with magnetic helicity is still a notable question \citep{Georgoulis09,wangcy15,Kim17,Park21}.  
This involves the basic issues of non-potential magnetic field and the relationship with magnetic reconnection, which is beyond the scope of discussion here.
 
 \item  Some basic problems related to observing and studying solar magnetic (current) helicity with the solar cycles. They involve the measurement and analysis accuracy of magnetic helicity and the completeness of data. It can be found from Eq. (\ref{eq:meancurr}) that it is impossible to obtain the complete current helicity value in the solar active regions, and it is also difficult to obtain complete data of vector magnetic field in all of the solar active regions on the solar surface. In the calculation of magnetic helicity transfer in the solar surface (such as Eq. (\ref{eq:heli35})), there are also problems in the measurement accuracy of magnetic field and velocity field, and it is difficult to cover the whole surface of the Sun. If we also consider the observation error of the instruments, it is naturally easy to understand that there are many uncertainties in the research, and the differences in the observation and calculation results of   different authors \citep[cf.] []{Ai89,Ronan92,WangHM92,Sakuari95,Sakuari01,ZhangHQ03,Pevtsov05,Xu12}. There are still some open questions until now.

\end{enumerate}

\section{Acknowledgements}

The authors would like to thank the Huairou Solar Observing Station, National Astronomical Observatories, Chinese Academy of Sciences, and {the Solar \& Heliospheric Observatory (SOHO)   for providing the observational data. The authors would like especially to thank Profs. D. Sokoloff, A. Brandenburg, T. Sakurai, S. Bao, Drs. Y. Gao, K. Kuzanyan, and V. Pipin for their kindly discussions in the study of the magnetic helicity with the solar observations.
This study is supported by grants from the National Natural Science Foundation (NNSF) of China under the project grant 11673033, 11427803, 11427901, 12073040, 12073041, 12003051, 11973056, and other grants at Huairou Solar Observing Station, National Astronomical Observatories, Chinese Academy of Sciences.S. Y. acknowledges support by grants 11427901, 10921303, 11673033, U1731113, 11611530679,
and 11573037 of the National Natural Science Foundation of China and grants no. XDB09040200, XDA04061002, XDA15010700 of
the Strategic Priority Research Program of the Chinese Academy of Sciences and the Youth Innovation Promotion Association of CAS (2019059).

\bibliography{wiley}%

\latexprintindex

\end{document}